\documentclass[tighten,nolinenumbers, notrackchanges, twocolumn,twocolappendix]{aastex631}

\usepackage{amsmath}
\usepackage{amssymb}
\usepackage{amssymb}
\usepackage{mathtools}
\usepackage{mathrsfs}

\usepackage{morefloats}
\usepackage{longtable}
\usepackage{afterpage}

\hypersetup{linkcolor=red,citecolor=blue,urlcolor=magenta}

\usepackage{soul}

\usepackage[makeroom]{cancel}

\graphicspath{{figs/}}

\usepackage{xspace}

\newcommand*{\XPSI}{X-PSI\xspace}
\newcommand*{\NICER}{NICER\xspace}
\newcommand*{\xmm}{XMM-Newton\xspace}
\newcommand*{\PyMultiNest}{\textsc{PyMultiNest}\xspace}
\newcommand*{\MultiNest}{\textsc{MultiNest}\xspace}

\newcommand{\msol}{$M_\odot$\xspace}
\newcommand{\jdbl}{PSR~J0030$+$0451\xspace}
\newcommand{\joh}{PSR~J0740$+$6620\xspace}

\newcommand{\TT}[1]{\texttt{#1}}

\newcommand{\be}{\begin{equation}}
\newcommand{\ee}{\end{equation}}

\defcitealias{Riley2021}{R21}
\defcitealias{Wolff20}{W21}
\defcitealias{Miller2021}{M21}
\defcitealias{Salmi2022}{S22}
\defcitealias{Dittmann2024}{D24}

\received{24 January 2024}
\revised{1 April 2024}
\revised{2 May 2024}
\revised{29 May 2024}
\accepted{31 May 2024}
\published{18 October 2024}

\submitjournal{The Astrophysical Journal}

\shorttitle{Radius of PSR J0740+6620}
\shortauthors{Salmi~et~al.}

\begin{document}

\title{The Radius of the High-mass Pulsar PSR J0740+6620 with 3.6 yr of NICER Data
}

\correspondingauthor{T.~Salmi}
\email{t.h.j.salmi@uva.nl}
  
\author[0000-0001-6356-125X ]{Tuomo~Salmi}
\affil{Anton Pannekoek Institute for Astronomy, University of Amsterdam, Science Park 904, 1098XH Amsterdam, the Netherlands}

\author[0000-0002-2651-5286]{Devarshi~Choudhury}
\affil{Anton Pannekoek Institute for Astronomy, University of Amsterdam, Science Park 904, 1098XH Amsterdam, the Netherlands}

\author[0000-0002-0428-8430]{Yves~Kini}
\affil{Anton Pannekoek Institute for Astronomy, University of Amsterdam, Science Park 904, 1098XH Amsterdam, the Netherlands}

\author[0000-0001-9313-0493]{Thomas~E.~Riley}
\affil{Anton Pannekoek Institute for Astronomy, University of Amsterdam, Science Park 904, 1098XH Amsterdam, the Netherlands}

\author[0000-0003-3068-6974]{Serena~Vinciguerra}
\affil{Anton Pannekoek Institute for Astronomy, University of Amsterdam, Science Park 904, 1098XH Amsterdam, the Netherlands}

\author[0000-0002-1009-2354]{Anna~L.~Watts}
\affil{Anton Pannekoek Institute for Astronomy, University of Amsterdam, Science Park 904, 1098XH Amsterdam, the Netherlands}

\author[0000-0002-4013-5650]{Michael~T.~Wolff}
\affil{Space Science Division, U.S. Naval Research Laboratory, Washington, DC 20375, USA}

\author[0009-0008-6187-8753]{Zaven~Arzoumanian}
\affil{X-Ray Astrophysics Laboratory, NASA Goddard Space Flight Center, Code 662, Greenbelt, MD 20771, USA}

\author[0000-0002-9870-2742]{Slavko~Bogdanov} 
\affil{Columbia Astrophysics Laboratory, Columbia University, 550 West 120th Street, New York, NY 10027, USA}

\author[0000-0001-8804-8946]{Deepto~Chakrabarty} 
\affil{Massachusetts Institute of Technology, Cambridge, MA, USA}

\author[0000-0001-7115-2819]{Keith~Gendreau}
\affil{X-Ray Astrophysics Laboratory, NASA Goddard Space Flight Center, Code 662, Greenbelt, MD 20771, USA}

\author[0000-0002-6449-106X]{Sebastien~Guillot}
\affil{Institut de Recherche en Astrophysique et Plan\'{e}tologie, UPS-OMP, CNRS, CNES, 9 avenue du Colonel Roche, BP 44346, F-31028 Toulouse Cedex 4, France}

\author[0000-0002-6089-6836]{Wynn~C.~G.~Ho}
\affil{Department of Physics and Astronomy, Haverford College, 370 Lancaster Avenue, Haverford, PA 19041, USA}

\author[0000-0002-1169-7486]{Daniela~Huppenkothen}
\affil{SRON Netherlands Institute for Space Research, Niels Bohrlaan 4, 2333 CA Leiden, the Netherlands}

\author[0000-0002-8961-939X]{Renee~M.~Ludlam}
\affil{Department of Physics and Astronomy, Wayne State University, 666 W Hancock, Detroit, MI 48201, USA}

\author[0000-0003-4357-0575]{Sharon~M.~Morsink}
\affil{Department of Physics, University of Alberta, 4-183 CCIS, Edmonton, AB, T6G 2E1, Canada}

\author[0000-0002-5297-5278]{Paul~S.~Ray}
\affil{Space Science Division, U.S. Naval Research Laboratory, Washington, DC 20375, USA}

\begin{abstract}

We report an updated analysis of the radius, mass, and heated surface regions of the massive pulsar PSR J0740+6620 using Neutron Star Interior Composition Explorer (NICER) data from 2018 September 21 to 2022 April 21, a substantial increase in data set size compared to previous analyses. 
Using a tight mass prior from radio timing measurements and jointly modeling the new NICER data with XMM-Newton data, the inferred equatorial radius and gravitational mass are $12.49_{-0.88}^{+1.28}$ km and $2.073_{-0.069}^{+0.069}$ \msol respectively, each reported as the posterior credible interval bounded by the $16\,\%$ and $84\,\%$ quantiles, with an estimated systematic error $\lesssim 0.1$ km. 
This result was obtained using the best computationally feasible sampler settings providing a strong radius lower limit but a slightly more uncertain radius upper limit. 
The inferred radius interval is also close to the $R=12.76_{-1.02}^{+1.49}$ km obtained by Dittmann et al., when they require the radius to be less than $16$ km as we do.
The results continue to disfavor very soft equations of state for dense matter, with $R<11.15$ km for this high-mass pulsar excluded at the 95\% probability. 
The results do not depend significantly on the assumed cross-calibration uncertainty between NICER and XMM-Newton.
Using simulated data that resemble the actual observations, we also show that our pipeline is capable of recovering parameters for the inferred models reported in this paper.
\end{abstract}

\section{Introduction}
\label{sec:intro}

Determination of the masses and radii of a set of neutron stars (NSs) can be used to infer properties of the high-density matter in their cores. 
This is possible due to the one-to-one mapping between the equation of state (EOS) and the mass--radius dependence of the NS \citep[see, e.g.,][]{LP2016,Baym2018}.
One way to infer mass and radius is to model the X-ray pulses produced by hot regions on the surface of a rapidly rotating NS including relativistic effects \citep[see, e.g.,][and references therein]{Watts2016,Bogdanov2019b}. 
For example, this technique has been applied in analyzing the data from NASA’s Neutron Star Interior Composition Explorer \citep[\NICER;][]{Gendreau2016} for rotation-powered millisecond pulsars. 
Their thermal emission is dominated by surface regions heated by the bombardment of charged particles from a magnetospheric return current \citep[see, e.g.,][]{RudermanSutherland1975,arons81,HM01}.
Results for two sources have been released (\citealt{MLD_nicer19,Riley2019,Miller2021,Riley2021,Salmi2022}, hereafter M21, R21, and S22, respectively; \citealt{Vinciguerra2024bravo}), providing useful constraints for dense matter models (see, e.g., \citetalias{Miller2021}; \citealt{Raaijmakers2021,Biswas2022,Annala2023,Takatsy23}).  
These results have also triggered studies on the magnetic field geometries and how nonantipodal they can be \citep[see, e.g.,][]{Bilous_2019,Chen2020,Kalapotharakos2021,Carrasco2023}. 

In this work, we use a new \NICER data set (with increased exposure time) to analyze the high-mass pulsar \joh, previously studied in \citetalias{Miller2021}, \citetalias{Riley2021}, and \citetalias{Salmi2022}.
In those works, the NS mass had a tight prior from radio timing ($2.08 \pm 0.07$ \msol; \citealt{Fonseca20}), and the NS radius was inferred to be, using both \NICER and \xmm data, $12.39_{-0.98}^{+1.30}$ km in \citetalias{Riley2021} and $13.7_{-1.5}^{+2.6}$ km in \citetalias{Miller2021}. 
However, the results were slightly sensitive to the inclusion of the \xmm data (used to better constrain the phase-averaged source spectrum and hence -- indirectly -- the NICER background) and assumptions made in the cross-calibration between the two instruments. 
In \citetalias{Salmi2022}, the use of \NICER background estimates \citep[such as ``3C50" from][]{remi22} was shown to yield results similar to the joint \NICER and \xmm analysis, giving confidence in the use of \xmm data as an indirect method of background constraint. 

In this paper we use a new \NICER data set with more than $1$ Ms additional exposure time and more than $0.5$ million additional observed counts, an $\sim 90\%$ increase in the counts, compared to the data sets used in \citetalias{Miller2021} and \citetalias{Riley2021}.
This is expected to reduce the uncertainties in the inferred NS parameters \citep{lomiller13,Psaltis2014}, and we explore whether this is indeed the case. 
We also look in detail at the influence of sampler settings on the credible intervals.

The remainder of this paper is structured as follows. 
In Section \ref{sec:data_and_bkg}, we introduce the new data set used for \joh.
In Section \ref{sec:methods}, we summarize the modeling procedure, and in Section \ref{sec:results} we present the results for the updated analysis. 
We discuss the implications of the results in Section \ref{sec:discussion} and conclude in Section \ref{sec:conclusions}.
Inference results using simulated data, resembling our new data, are shown in Appendix \ref{sec:simulations}.
\setcounter{footnote}{10}
{
    \begin{figure}[t!]
    \centering
    \resizebox{\hsize}{!}{\includegraphics[
    width=\textwidth]{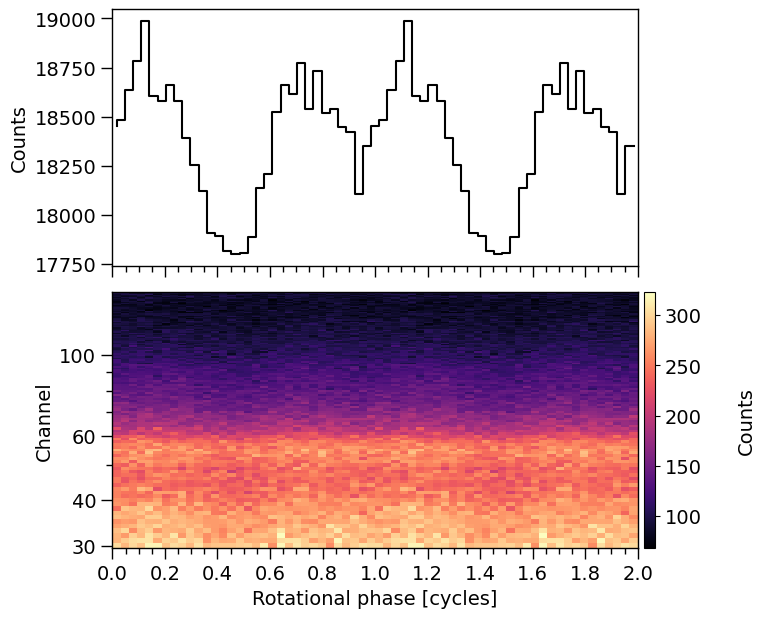}}
    \caption{\small{
The new phase-folded \joh event data for two rotational cycles (for clarity).
The top panel shows the pulse profile summed over the channels. 
As in Figure 1 of \citetalias{Riley2021}, the total number of counts is given by the sum over all phase--channel pairs (over both cycles).
For the modeling all the event data are grouped into a single rotational cycle instead.
    }}
    \label{fig:data_gamma}
    \end{figure}
}

\section{Modeling Procedure}\label{sec:methods}

The modeling procedure is largely shared with that of \citet{Bogdanov2019b}, \citet{Riley2019}, \citet{Bogdanov2021}, \citetalias{Riley2021}, and \citetalias{Salmi2022}. 
We use the X-ray Pulse Simulation and Inference\footnote{\url{https://github.com/xpsi-group/xpsi}} (\XPSI; \citealt{xpsi}) code, with versions ranging from \texttt{v0.7.10} to \texttt{v1.2.1}\footnote{The versions are practically identical for the considered models; the only actual difference is the fix of a numerical ray tracing issue since \texttt{v0.7.12}, affecting only a few parameter vectors with emission angles extremely close to $90\degr$ (https://github.com/xpsi-group/xpsi/issues/53). This is not expected to alter the inferred radius.} for inference runs (\texttt{v1.2.1} used for the headline results), and \texttt{v2.2.1} for producing the figures.
Complete information of each run, including the exact \XPSI version, data products, posterior sample files, and all the analysis files can be found in the Zenodo repository at 
doi:\href{https:/doi.org/10.5281/zenodo.10519473}{10.5281/zenodo.10519473}.
%of \citet{salmi_zenodo24gamma}. 
In the next sections we summarize the modeling procedure and focus on how it differs from that used in previous work.

\subsection{X-Ray Event Data}\label{sec:data_and_bkg}

The \NICER X-ray event data used in this work were processed with a similar procedure as the previous data reported in \citet{Wolff21} and used in \citetalias{Miller2021} and \citetalias{Riley2021}, but with some notable differences (note that a completely different 3C50 procedure was applied in \citetalias{Salmi2022}).
The new data were collected from a sequence of exposures, in the period 2018 September 21 $-$ 2022 April 21 (observation IDs, hereafter obsIDs, 1031020101 through 5031020445), whereas the period of the previous data set began on the same start date but ended on 2020 April 17 (using the obsIDs shown in \citealt{Wolff21}).
After filtering the data (described below), this resulted in 2.73381 Ms of on-source exposure time, compared to the previous 1.60268 Ms.

The filtering procedure differed slightly from that used in earlier work. 
First, similarly to the previous work, we rejected data obtained at low cut-off rigidities of the Earth magnetic field (COR\_SAX $<2~\mathrm{GeV} c^{-1}$ \footnote{Note this number was reported incorrectly in \citetalias{Miller2021} and \citetalias{Riley2021} but this had no effect on the outcome of the analysis.}) to minimize high-energy particle interactions indistinguishable from X-ray events, and we excluded all the events from the noisy detector DetID 34 and the events from DetID 14 when it had a count rate greater than 1.0 counts per second (cps) in 8.0 s bins.
We also cut all 2 s bins with total count rate larger than 6 cps to remove generally noisy time intervals.
However, here the previously used sorting method of good time intervals (GTIs) from \citet{Guillot2019} and filtering based on the angle between \joh and the Sun were not applied.
Instead, a maximum ``undershoot rate" (i.e., detector reset rate) of 100 cps per detector was imposed to produce a cleaned event list with less contamination from the accumulation of solar optical photons (``optical loading," up to $\sim 0.4$ keV) typically---but not exclusively---happening at low Sun angles. 
A test of the event extraction procedure holding all other criteria constant and just varying the maximum undershoot rate from a value of 50 cps to 200 cps (in increments of 50 cps) gave us four different filtered event lists from the same basic list of obsIDs. 
Testing each event list for pulsation detection significance using a Z$_2^2$ test \citep{Buccheri1983} showed that the pulsar was clearly detected at highest significance for the maximum undershoot rate of 100 cps and thus it is this event list that we settled on for this analysis.

We note that our new procedure is expected to be less prone to the type of systematic selection bias suggested for the older GTI sorting method in \citet[][although even there the effect was only marginal as discussed in Section 2.1 of \citetalias{Salmi2022}]{Essick2022}.
This is because the undershoot rates do not correlate with the flux from the optically dim pulsar and the detection significance is maximized only by selecting from four different options.

In the pulse profile analysis, we used again the pulse invariant (PI) channel subset [30,150), corresponding to the nominal photon energy range [0.3, 1.5] keV, as in \citetalias{Riley2021} and \citetalias{Salmi2022}, unless mentioned otherwise.
The number of rotational phase bins in the data is also 32 as before.
The data split over two rotational cycles are visualized in Figure \ref{fig:data_gamma}.

For the analyses including \xmm observations, we used the same phase-averaged spectral data and blank-sky observations (for background constraints) as in \citetalias{Miller2021}, \citetalias{Riley2021}, and \citetalias{Salmi2022} with the three EPIC instruments (pn, MOS1, and MOS2).
Unless mentioned otherwise, we included the energy channels [57,299) for pn, and [20,100) for both MOS1 and MOS2, which are the same choices as in \citetalias{Riley2021} and \citetalias{Salmi2022} (although this was not explicitly stated in those papers). \citetalias{Miller2021} made the same choices except for including one more high-energy channel for the pn instrument. 
The data are visualized in Figures 4 and 16 of \citetalias{Riley2021}.

\subsection{Instrument Response Models}\label{sec:response}

For the \xmm EPIC instruments we used the same ancillary response files (ARFs) and 
redistribution matrix files (RMFs) as in \citetalias{Riley2021} 
and \citetalias{Salmi2022}. 
For the \NICER events utilized in this study, the calibration version is \texttt{xti20210707}, the HEASOFT version is \texttt{6.30.1} containing \texttt{NICERDAS 2022-01-17V009}. 
We generated response matrices differently from the previous analyses in that 
we used a combined response file,\footnote{Defined as a product of ARF 
and RMF, which were created through the \texttt{nicerarf} and \texttt{nicerrmf} tools.} which was tailored for the focal plane module (FPM) information now 
maintained in the NICER FITS event files. 
Thus, for the analysis in this study, the effective area will reflect the exact 
FPM exposures resulting in the rescaling of effective area to account for 
the complete removal of one detector (DetID 34) and the partial removal of 
another detector (DetID 14) as addressed above. 
The resulting calibration products are available on Zenodo at doi:\href{https:/doi.org/10.5281/zenodo.10519473}{10.5281/zenodo.10519473}.
%\citep{salmi_zenodo24gamma}.

When modeling the observed signal, we allowed uncertainty in the effective areas of both \NICER and all three \xmm detectors, due the lack of an absolute calibration source. 
As in \citetalias{Riley2021} and \citetalias{Salmi2022}, we define energy-independent effective area scaling factors as:
\begin{equation}
\alpha_{\mathrm{NICER}} = \alpha_{\mathrm{SH}}\alpha'_{\mathrm{NICER}} \text{ and }
\alpha_{\mathrm{XMM}} = \alpha_{\mathrm{SH}}\alpha'_{\mathrm{XMM}},
\label{eq:scaling_factors}
\end{equation}
where $\alpha_{\mathrm{NICER}}$ and $\alpha_{\mathrm{XMM}}$ are the overall scaling factors for \NICER and \xmm, respectively (used to multiply the effective area of the instrument), $\alpha_{\mathrm{SH}}$ is a shared scaling factor between all the instruments (to simulate absolute uncertainty of X-ray flux
calibration), and $\alpha'_{\mathrm{NICER}}$ and $\alpha'_{\mathrm{XMM}}$ are telescope-specific scaling factors (to simulate relative uncertainty between the \NICER and \xmm calibration). 
As before, we assume that the factors are identical for pn, MOS1, and MOS2 (which may not be true). 
In our headline results, we apply the restricted 10.4\% uncertainty  \citep{Ishida2011,Madsen2017,Plucinsky2017}\footnote{See also \url{https://xmmweb.esac.esa.int/docs/documents/CAL-TN-0018.pdf}.} in the overall scaling factors, as in the exploratory analysis of Section 4.2 in \citetalias{Riley2021} (see also Section 3.3 of \citetalias{Salmi2022}), which results from assuming a $10\,\%$ uncertainty in $\alpha_{\mathrm{SH}}$ and a $3\,\%$ uncertainty in the telescope-specific factors.
This choice is tighter than the 15\% uncertainty used in the main results of \citetalias{Riley2021} (who assumed $10.6\,\%$ uncertainty in both shared and telescope-specific factors), but is similar to that used in \citetalias{Miller2021} and \citetalias{Salmi2022}. 
However we have also explored the effect of different choices for the effective-area scaling factors in Section \ref{sec:nicerXxmm}.

\subsection{Pulse Profile Modeling Using X-PSI}\label{sec:pp_modeling}

As in previous \NICER analyses \citepalias[e.g.,][]{Miller2021,Riley2021,Salmi2022}, we use the `Oblate Schwarzschild' approximation to model the energy-resolved X-ray pulses from the NS \citep[see, e.g.,][]{ML98,NS02,PG03,CMLC07,MLC07,lomiller13,AGM14,Bogdanov2019b,Watts2019}. 
In addition, we use now a corrected BACK\_SCAL factor \citepalias[see Section 3.4 of][]{Salmi2022}. 
As before, we use the mass, inclination, and distance priors from \citet{Fonseca20}, interstellar attenuation model \texttt{TBabs} \citep[updated in 2016]{Wilms2000}, and the fully ionized hydrogen atmosphere model \texttt{NSX} \citep{Ho01}.
As shown in \citet{Salmi2023}, the assumptions for the atmosphere do not seem to significantly affect the inferred radius of \joh. 
A larger sensitivity to the atmosphere choices was found in \citet{Dittmann2024}, but this could be related to their sampling methodology or their  larger prior space, e.g., including NS radii above 16 km.
The shapes of the hot emitting regions are again characterized using the single-temperature-unshared (\texttt{ST-U}) model with two circular uniform temperature regions \citep{Riley2019}.

\subsection{Posterior Computation}\label{sec:posterior_computation}

We compute the posterior samples using \PyMultiNest \citep{PyMultiNest} and \MultiNest \citep{MultiNest_2008,multinest09,FHCP2019}, as in the previous \XPSI analyses.
For the headline joint \NICER and \xmm results of this work, we use the following \MultiNest settings: $4\times10^{4}$ live points and a $0.01$ sampling efficiency (SE).\footnote{Note that this factor is not the nominal \MultiNest SE parameter (SE'), since SE' is modified in \XPSI to account for the initial prior volume that is smaller than unity as explained in \citet{riley_thesis} and in Appendix \ref{sec:sampling_efficiency}. 
The SE' value is about $6.9$ times higher than the SE value for all the models in this paper.
\label{footnote:SE}}
For the NICER-only analysis, and the exploratory NICER-XMM analysis, we used the same number of live points but SE = 0.1.  We discuss sensitivity to sampler settings further in Section \ref{sec:results}.

{
    \begin{figure*}[!htbp]
    \centering
    \includegraphics[
    width=0.49\textwidth]
    {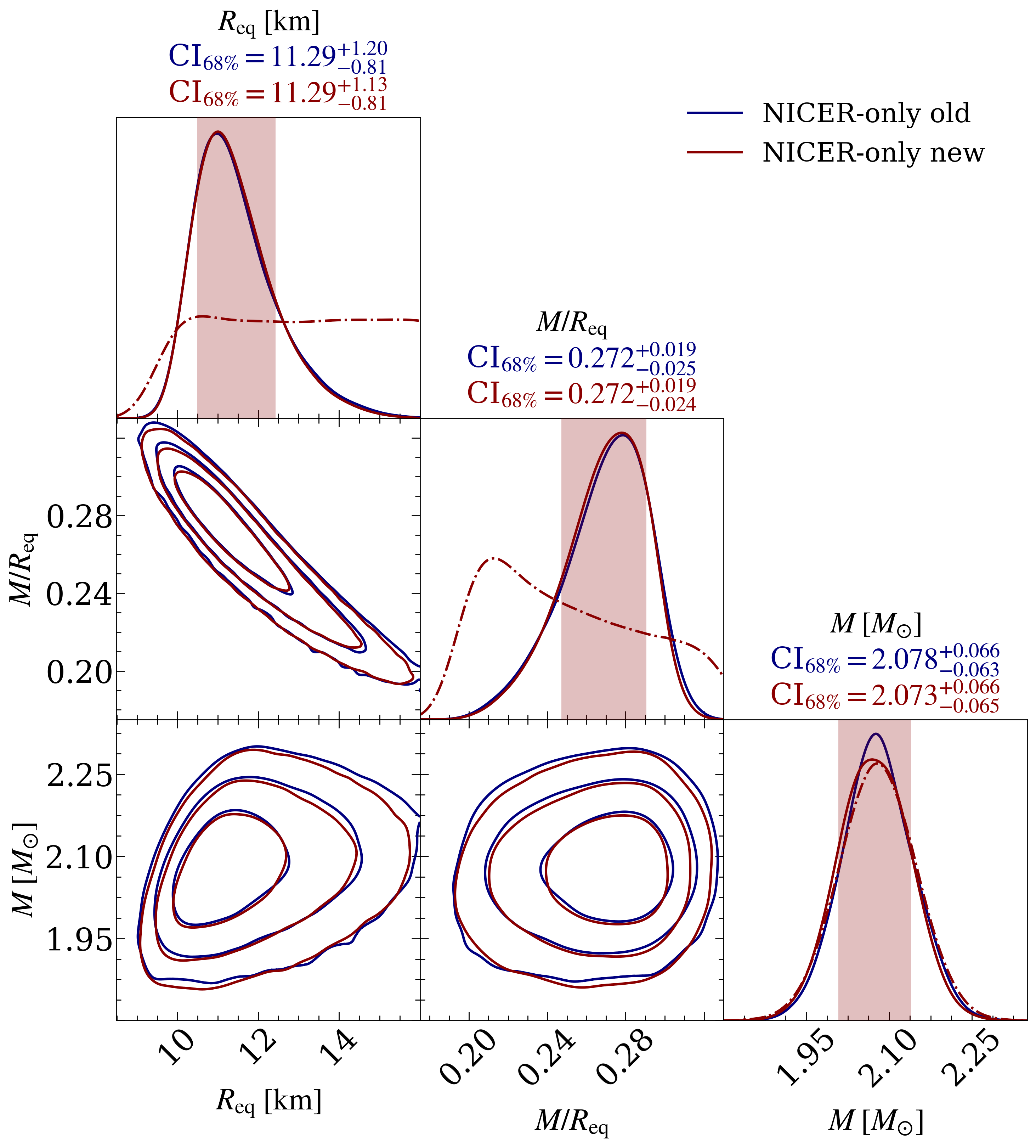}
    \includegraphics[
    width=0.49\textwidth]
    {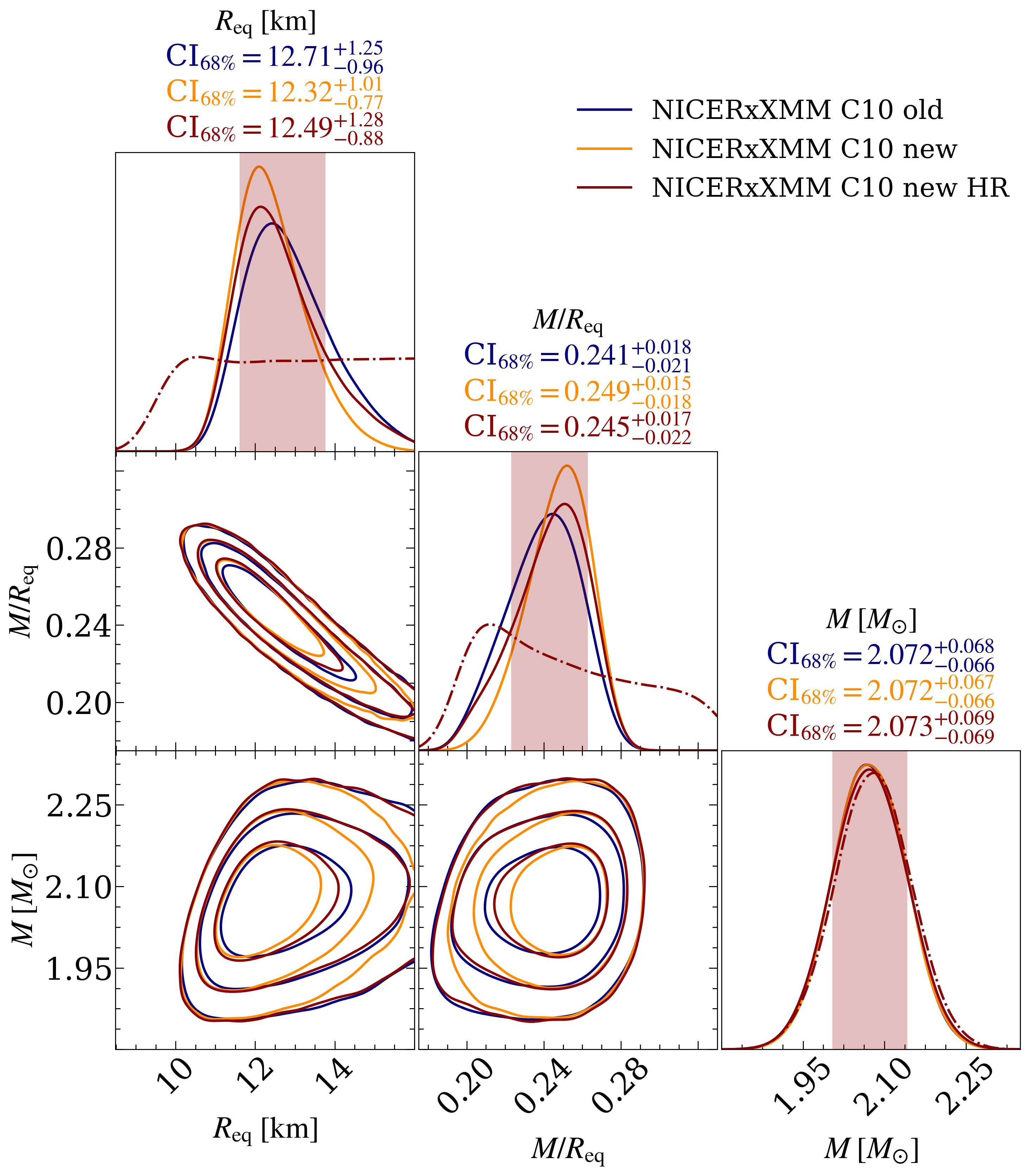}  
    \caption{\small{
    Radius, compactness, and mass posterior distributions using the new \NICER data set and \texttt{ST-U} model in the \NICER-only analysis (left panel) and in the joint \NICER and \xmm analysis (right panel) compared to the old results from \citetalias{Riley2021}.
    Here “C10” refers to a $\pm 10.4~\%$ calibration uncertainty in the overall effective-area scaling factors, ``new" and ``old" without qualification have SE = 0.1 and ``HR" refers to the new headline results with SE = 0.01.
    Dash--dotted functions represent the marginal prior probability density functions (PDFs). 
    The shaded vertical bands show the $68.3\%$ credible intervals (for the posteriors corresponding to the red curves), and the contours in the off-diagonal panels show the $68.3\%$, $95.4\%$, and $99.7\%$ credible regions. 
    See the captions of Figure 5 of \citetalias{Salmi2022} and Figure 5 of \citetalias{Riley2021} for additional details about the figure elements.
    }}
    \label{fig:mr_posteriors}
    \end{figure*}
}

%\figsetstart
%\figsetnum{3}
%\figsettitle{Posteriors other than space-time parameters}

%\figsetgrpstart
%\figsetgrpnum{3.1}
%\figsetgrptitle{NICERxXMM geometry parameters}
%\figsetplot{f3_1.png}
%\figsetgrpnote{Posterior distributions for the geometry parameters in the joint \NICER and \xmm analysis. See the caption in the main text for more details.}
%\figsetgrpend

%\figsetgrpstart
%\figsetgrpnum{3.2}
%\figsetgrptitle{NICER-only geometry parameters}
%\figsetplot{f3_2.png}
%\figsetgrpnote{Posterior distributions for the geometry parameters in the \NICER-only analysis. See the caption in the main text for more details.}
%\figsetgrpend

%\figsetgrpstart
%\figsetgrpnum{3.3}
%\figsetgrptitle{NICERxXMM other parameters}
%\figsetplot{f3_3.png}
%\figsetgrpnote{Posterior distributions for the remaining parameters in the joint \NICER and \xmm analysis. See the caption in the main text for more details.}
%\figsetgrpend

%\figsetgrpstart
%\figsetgrpnum{3.4}
%\figsetgrptitle{NICER-only other parameters}
%\figsetplot{f3_4.png}
%\figsetgrpnote{Posterior distributions for the remaining parameters in the \NICER-only analysis. See the caption in the main text for more details.}
%\figsetgrpend

%\figsetend

{
    \begin{figure*}[t!]
    \centering
    \includegraphics[
    width=\textwidth]
    {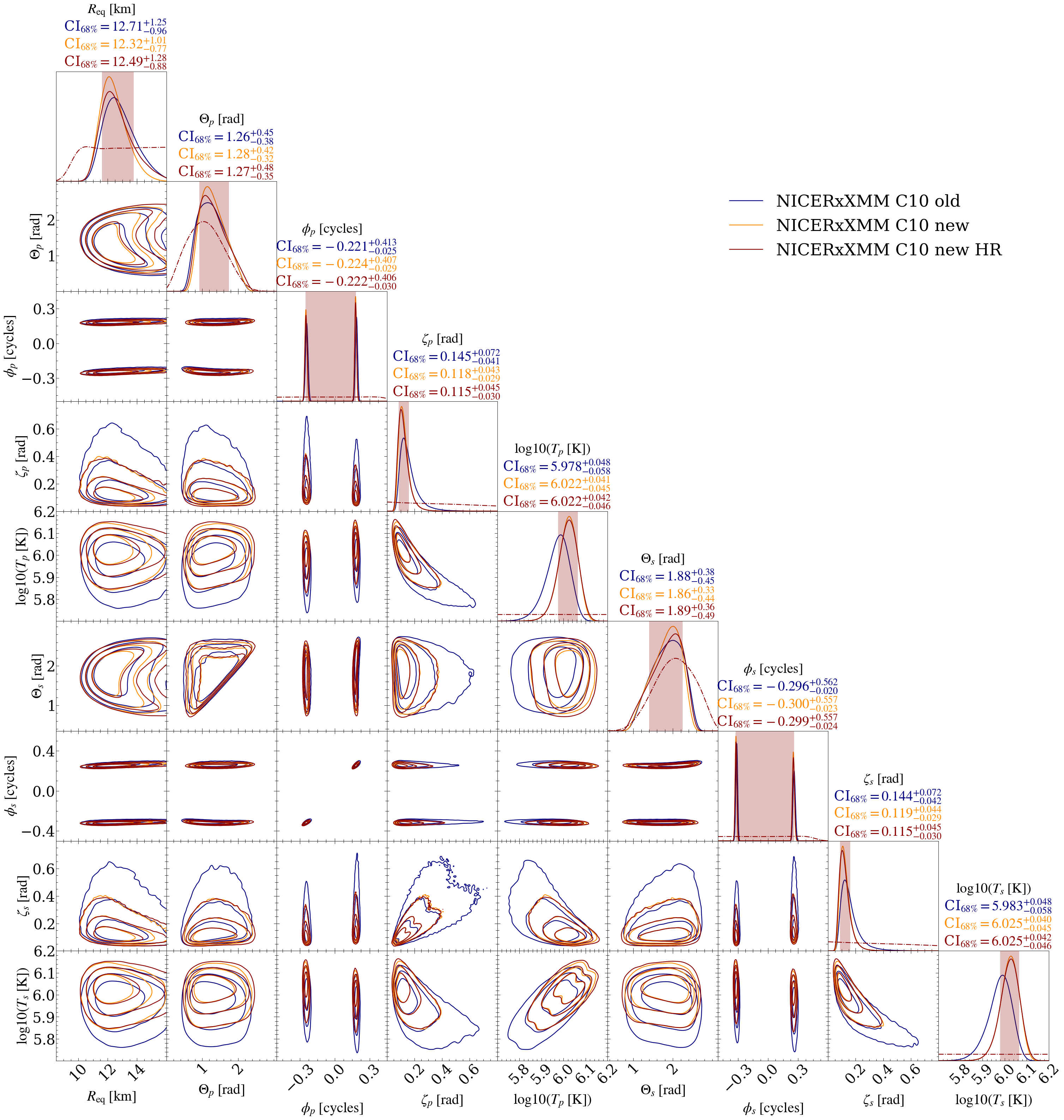}  
    \caption{\small{
    Posterior distributions for the hot region parameters using the new \NICER data set and \texttt{ST-U} model in the joint \NICER and \xmm analysis compared to the old results from \citetalias{Riley2021}. 
    See the captions of Figure \ref{fig:mr_posteriors} for more details about the figure elements.
    The complete figure set (four images), including the posterior distributions for the remaining parameters (both for \NICER-only and the joint \NICER and \xmm analyses) is available in the online journal (HTML version).
    }}
    \label{fig:other_posteriors}
    \end{figure*}
}

{
    \begin{figure*}[!htbp] %[t!]
    \centering
    \includegraphics[
    width=0.49\textwidth]
    {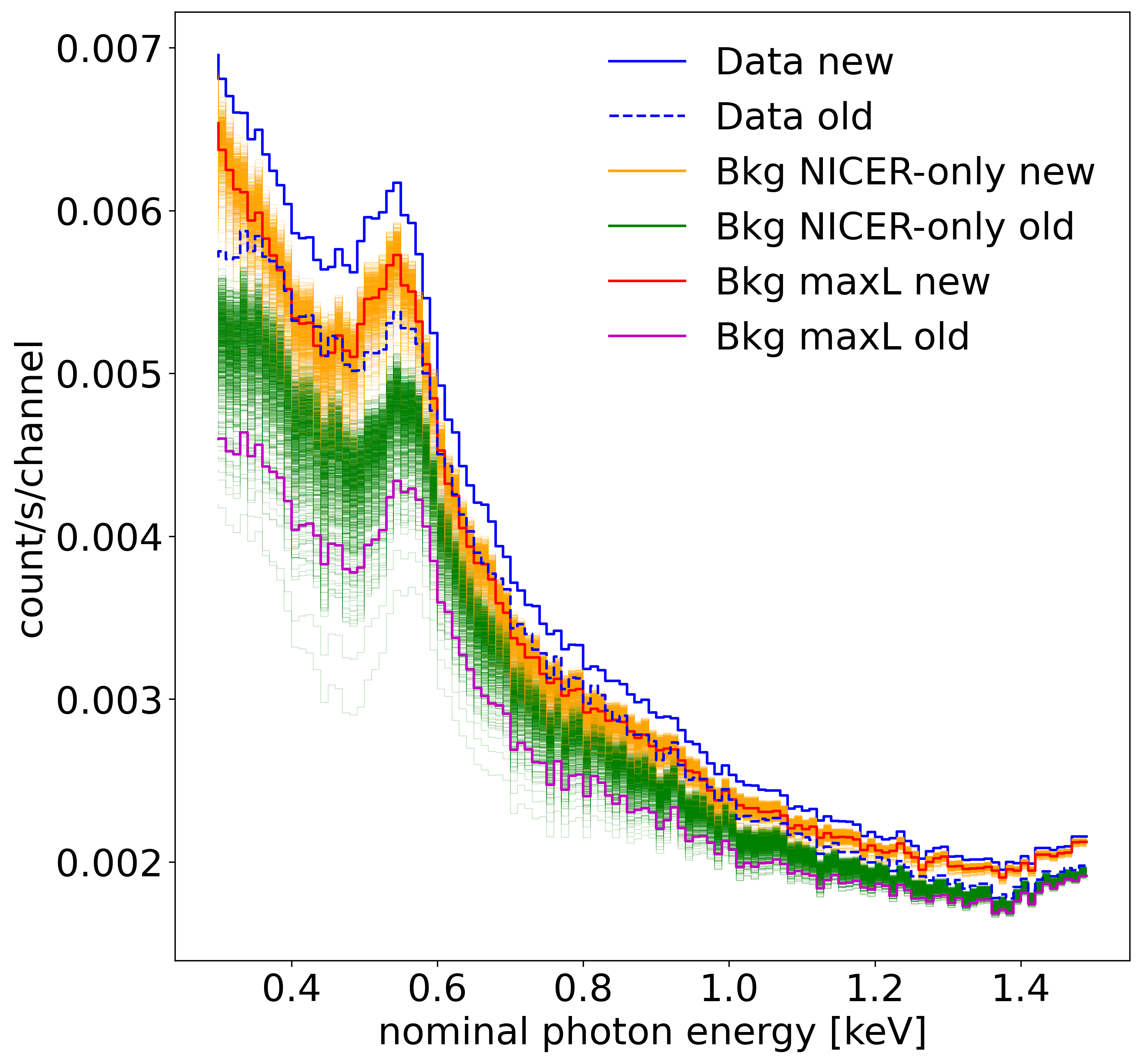}
    \includegraphics[
    width=0.49\textwidth]
    {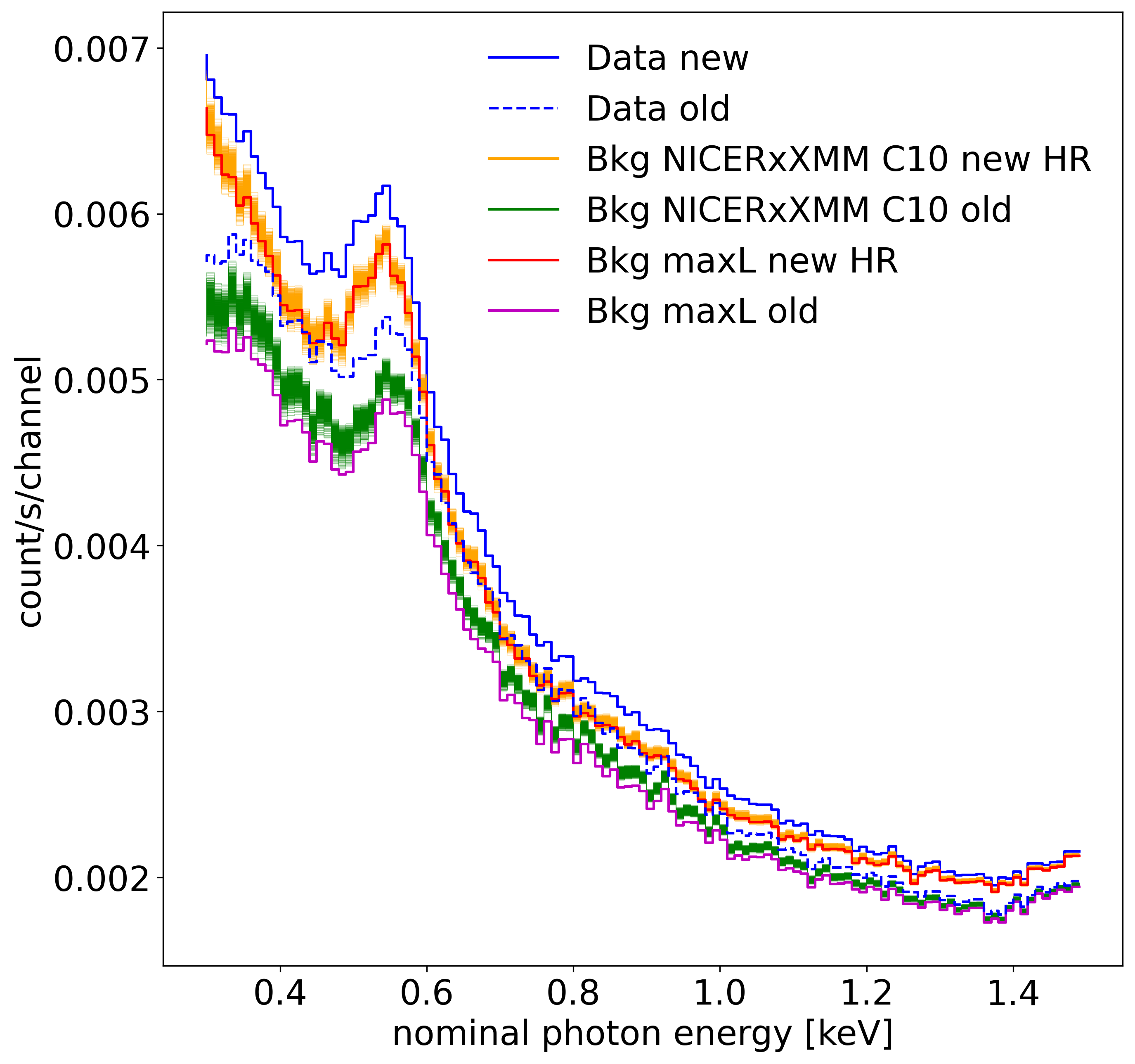}  
    \caption{\small{
    Comparison of the inferred \NICER background for different data sets and analyses. 
    Left panel: the blue solid and dashed stepped curves show the total \NICER count-rate spectra for the new and old data sets, respectively (note that the count rate has slightly changed because of the different filtering described in Section \ref{sec:data_and_bkg}). 
    The orange and green stepped curves show the background curves that maximize the likelihood for 1000 equally weighted posterior samples in the \NICER-only analysis with the new and old data, respectively.
    Accordingly, the red and magenta stepped curves show the background curves corresponding to the maximum likelihood sample for each run. 
    Right panel: same as the left panel, except the inferred backgrounds are now shown for the joint \NICER and \xmm runs with the new and old data, respectively (new results are shown for the HR run, but an almost identical background is inferred when using the old sampler settings)}.
    }
    \label{fig:bkg_inferred_data}
    \end{figure*}
}

{
    \begin{figure}[t!]
    \centering
    \resizebox{\hsize}{!}{\includegraphics[
    width=\textwidth]{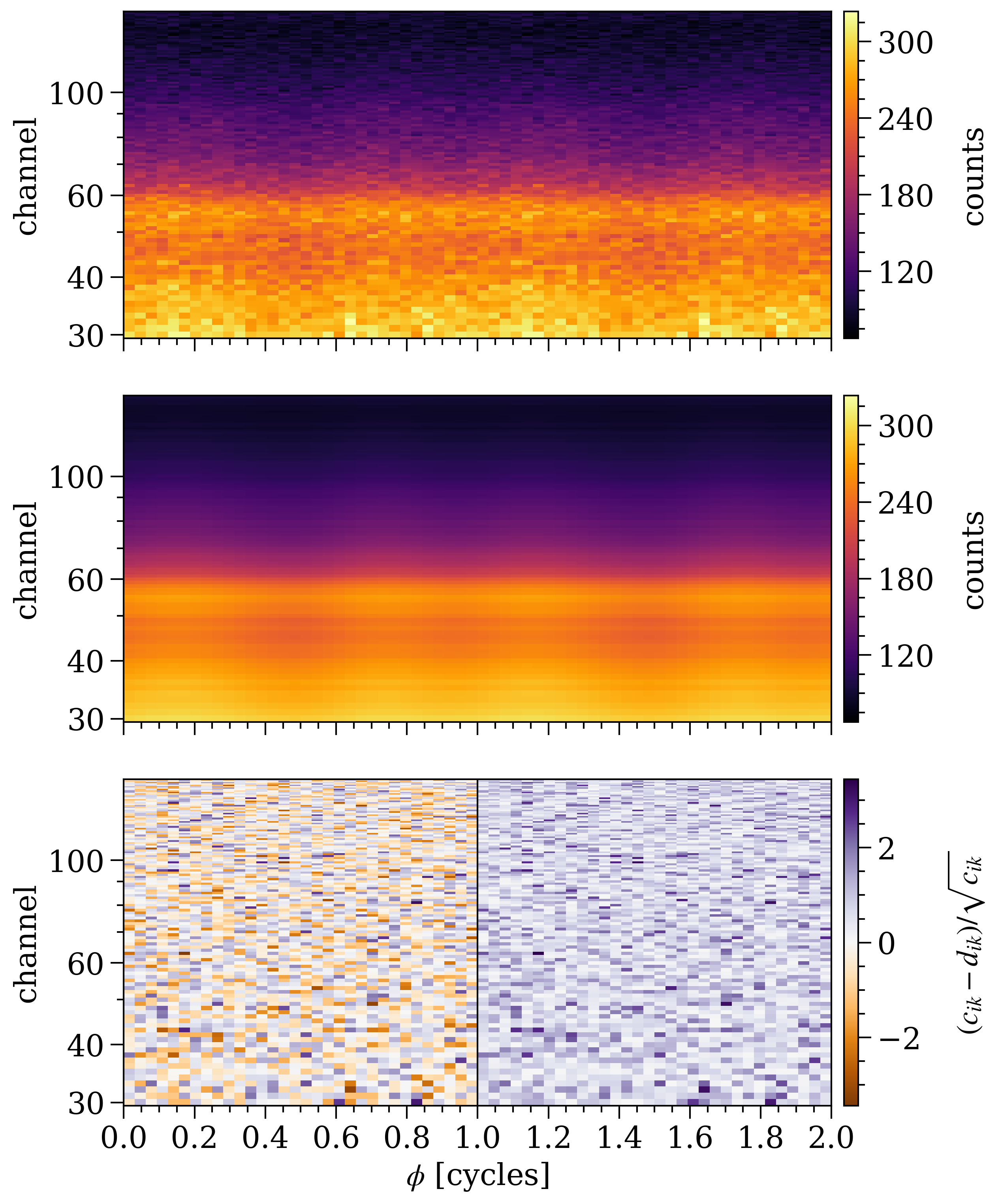}}
    \caption{\small{
The new \NICER count data, posterior-expected count numbers (averaged from 200 equally weighted posterior samples), and (Poisson) residuals for the \texttt{ST-U} model in the joint \NICER and \xmm analysis (for the HR run). 
See Figure 6 of \citetalias{Riley2021} for additional details about the figure elements.
    }}
    \label{fig:resid0}
    \end{figure}
}

{
    \begin{figure}[t!]
    \centering
    \resizebox{\hsize}{!}{\includegraphics[
    width=\textwidth]
    {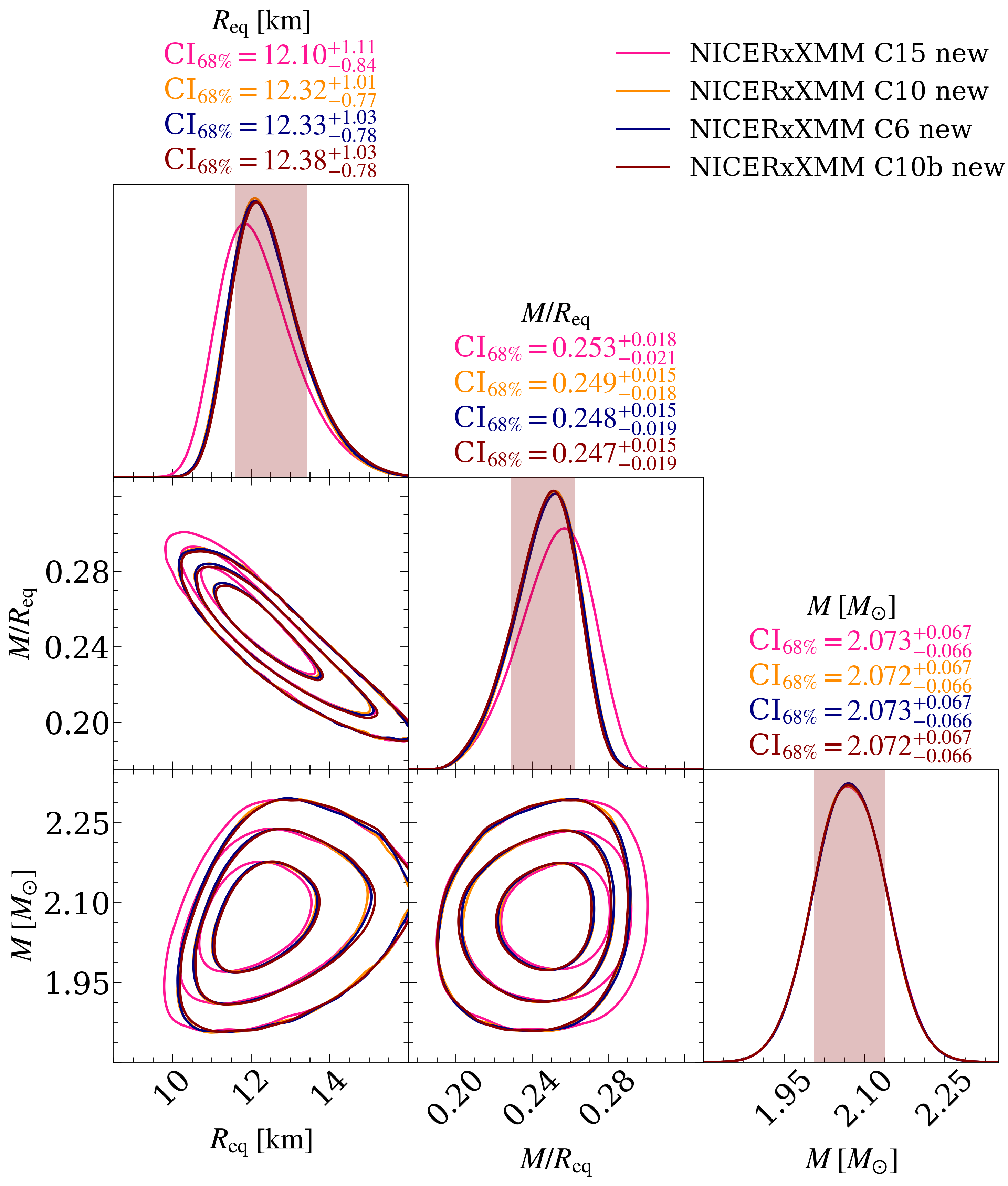}}
    \caption{\small{
     Posterior distributions for the space-time parameters using the new \NICER data set and \texttt{ST-U} model in the joint \NICER and \xmm analyses, with different assumptions for the effective-area scaling uncertainty and the used energy channels.
    Here “C15”, “C10”, and “C6” refer to runs with $\pm 15~\%$, $\pm 10.4~\%$, and $\pm 5.8~\%$ uncertainties in the overall effective-area scaling factors, respectively, and “C10b” to a run with a $\pm 10.4~\%$ uncertainty and an alternative channel choice (see the text at the end of Section \ref{sec:nicerXxmm}).
    The contours for the three latter cases are almost exactly overlapping. 
    See the caption of Figure \ref{fig:mr_posteriors} for additional details about the figure elements.  
    }}
    \label{fig:mr_posteriors_cross-calib}
    \end{figure}
}

\section{Inferences}\label{sec:results}

We start first by presenting the inference results for the updated \NICER-only analysis, and then proceed to the headline results obtained by fitting jointly the \NICER and \xmm data. 
To test the robustness of the analysis, the corresponding inference results using a synthetic data set are presented in Appendix \ref{sec:simulations}. 
In what follows we focus primarily on the constraints on radius since the posterior on the mass is essentially unchanged from the highly informative radio prior. 

\subsection{\NICER-only Fit}\label{sec:nicer-only}

When analyzing only the new \NICER data set (described in Section \ref{sec:data_and_bkg}), we find better constraints for some of the model parameters, but not for all of them.
As seen in the left panel of Figure \ref{fig:mr_posteriors}, no significant improvement in radius constraints is found compared to the old results from \citetalias{Riley2021} (where $R_{\textrm{eq}}=11.29_{-0.81}^{+1.20}$ km). 
These old results used a larger effective-area uncertainty and applied importance sampling instead of directly sampling with the most updated mass, distance, and inclination priors.\footnote{However, we checked that reanalyzing the old data with the new model choices does not significantly affect the results.} 
For the new data, the inferred radius is $11.29_{-0.81}^{+1.13}$ km and mass is $2.073_{-0.065}^{+0.066}$ \msol. 
However, a few of the other parameters, e.g., the sizes and temperatures of the emitting regions, are better constrained with the new data, as seen in the online images of Figure set \ref{fig:other_posteriors}.

The inferred background is also more tightly constrained and the source versus total count-rate ratio is significantly smaller for the new data 
(about $5-10\,\%$, in contrast to the previous $5-20\,\%$) as seen in the left panel of Figure \ref{fig:bkg_inferred_data}.  
This could explain why the radius constraints do not get tighter (see the discussion in Section \ref{sec:discussion}). 
In particular, the background corresponding to the maximum likelihood sample is larger for the new data (source versus total ratio is about $7\,\%$ instead of $17\,\%$). 
However, we note that with the old data most of the equally weighted posterior samples (and also the maximum posterior sample) have larger backgrounds than the maximum likelihood sample, which is shown in Figure \ref{fig:bkg_inferred_data} where the magenta curve is below most of the green curves (and maximum posterior source versus total ratio is as small as $5\,\%$). 
In contrast, the maximum likelihood sample for the new data has a background that is close to the average and maximum posterior samples.

\subsection{\NICER and \xmm Fit}\label{sec:nicerXxmm}

Analyzing the new \NICER data jointly with the \xmm data (right panel of Figure \ref{fig:mr_posteriors}), we find an inferred radius of $12.49_{-0.88}^{+1.28}$ km for our headline results (``new HR'' in the figure, see Section \ref{sec:posterior_computation}). 
For comparison we show the radius constraints obtained in \citetalias{Riley2021} with comparable cross-calibration uncertainty (``C10 old" here, Figure 14 of \citetalias{Riley2021}\footnote{Note that the old joint \NICER and \xmm results had an incorrect BACK\_SCAL factor and were obtained by importance sampling from a larger 15\% effective-area uncertainty to the 10.4\% uncertainty, rather than directly sampling with the 10.4\% uncertainty. 
However, these issues are not expected to have any significant effect on the results \citep[the BACK\_SCAL issue was also tested in][with a low-resolution run]{J0740zenodo_updated}.}), $12.71_{-0.96}^{+1.25}$ km, and the constraints we obtained for the new data but with the same sampler settings as in the older analysis (SE = 0.1), $12.32_{-0.77}^{+1.01}$ km. 
The new radius interval is thus about $\sim 20\%$ tighter (and shifted to smaller values) than the old when using the same sampler settings, but roughly equally tight when using the new settings.

The lower limit on the 68\% credible interval for the radius appears less sensitive to the sampler settings than the upper limit. 
Investigation of the likelihood surface reveals why this is the case. 
The likelihood falls off sharply at the lowest radii; for very small stars it is simply not possible to meet both the background and pulsed amplitude constraints given the tight geometric prior on the observer inclination for this source \citep{Fonseca20}. 
However the likelihood surface for radii above the maximum at $\sim 11$ km is very flat, making it harder to constrain the upper limit. 
In addition, as one approaches the highest radii, there are more solutions with hot spots that are smaller, hotter, and closer to the poles (see Figure \ref{fig:other_posteriors}). 
The likelihood surface in the space of spot size and temperature has a sharp point in this region of parameter space, and it is therefore important to resolve it well. 
In moving from SE = 0.1 to SE = 0.01 we observed that this region was sampled more extensively, and it appears to be this change that leads to the increase in the upper limit of the radius credible interval as SE gets smaller. 
The computational cost for the smaller SE, however, is much higher, and further increases in live points or reductions in SE to check convergence of the credible interval were not feasible. 

To check whether we had now mapped the likelihood surface in spot size--temperature space sufficiently well to determine exactly how the likelihood falls off, we performed an additional high-resolution (40,000 live points, SE = 0.01) run, but restricting the prior on the secondary spot temperature to log$_{10} T_s [K] > 6.15$. 
The sharp end of the likelihood surface was much more thoroughly sampled, with the drop-off in likelihood now well characterized. 
The inferred radius for the restricted prior run is $12.61^{+1.25}_{-0.87}$ km, but with overall maximum likelihoods lower than those in the full prior run. 
This suggests that any further increase of the upper limit of the radius credible interval in the full prior run (using more computational resources) is unlikely to be more than $\sim 0.1$ km. 
We take this as an estimate for the systematic error in the full prior run.

As in \citetalias{Riley2021} and \citetalias{Salmi2022}, the inferred radius for the joint \NICER and \xmm case is larger than for the \NICER-only case. 
However, this time the median values from the joint analyses are slightly closer to the \NICER-only result.
The new radius results are also slightly more constrained than in \citetalias{Salmi2022}, where the inferred radius was $12.90_{-0.97}^{+1.25}$ km using the 3C50-filtered \NICER data set (with a lower background limit) and \xmm data.
The inferred values for all of the parameters for the HR run are shown in Table \ref{table:results}, and the remaining posterior distributions are shown in Figure \ref{fig:other_posteriors}.
From there we see that, compared to the old \citetalias{Riley2021} results, consistent but slightly tighter constraints are obtained for hot region temperatures and sizes. The credible intervals for these parameters do not seem to depend significantly on the sampler settings used.
Even though the new data are more restrictive, we find that the \texttt{ST-U} model employed can still reproduce the data well, as seen from the residuals in Figure \ref{fig:resid0}.

Just as for the \NICER-only case (in Section \ref{sec:nicer-only}), the new inferred best-fitting \NICER background versus source count ratio changes compared to the previous analysis (see the right panel of Figure \ref{fig:bkg_inferred_data}).
However the change is smaller for the joint \NICER and \xmm analysis (regardless of the sampler settings); the maximum likelihood source count rate versus total count rate is now around $5.5\,\%$ instead of the previous $8.5\,\%$. 
When looking at the bulk of the equally weighted posterior samples, the difference is even smaller; 
in both cases the source count rate versus total count-rate range is around $4-7\,\%$.
The smaller change is entirely to be expected, since \xmm acts indirectly as a constraint on the \NICER background and already did so in the older analysis.

We also found (for exploratory analysis using SE = 0.1) that different assumptions for the cross-calibration scaling factors do not significantly affect the new results. 
This is shown in Figure \ref{fig:mr_posteriors_cross-calib}.
We see that the median radius only increases from around $12.1$ to around $12.3$ km when applying a $\pm 10.4\,\%$ instead of a $\pm 15\,\%$ uncertainty in the overall scaling factors. 
However, applying $\pm 5.8\,\%$ uncertainty\footnote{Which comes from assuming a $5\,\%$ uncertainty in the shared scaling factor and a $3\,\%$ uncertainty in the telescope-specific factors as in the test case of \citetalias{Salmi2022}.} produces almost identical results to those of $\pm 10.4~\%$.
We also explored the effect of a different choice for the energy channels included in the analysis.
As shown in Figure \ref{fig:mr_posteriors_cross-calib}, we found no significant difference in the inferred radius when using the channel choices of \citetalias{Miller2021} instead of those reported in Section \ref{sec:data_and_bkg} (i.e., \NICER channels only up to 123 instead of 149 and \xmm pn channels up to 299 instead of 298).

%\clearpage

\begin{deluxetable*}{lccccc}
\caption{Summary Table for the New Joint \NICER and \xmm Results}\label{table:results}
\tablehead{
\colhead{Parameter} & \colhead{Description} & \colhead{Prior PDF (density and support)} & \colhead{$\widehat{\textrm{CI}}_{68\%}$} & \colhead{$\widehat{D}_{\textrm{KL}}$} & \colhead{$\widehat{\textrm{ML}}$}
}
\startdata
$P$ $[$ms$]$ &
coordinate spin period &
$P=2.8857$, fixed &
$\cdots$ &
$\cdots$ &
$\cdots$ \\
\hline
$M$ $[M_{\odot}]$ &
gravitational mass &
$M, \cos(i)\sim N(\boldsymbol{\mu}^{\star},\boldsymbol{\Sigma}^{\star}) $ &
$2.073_{-0.069}^{+0.069}$ &
$0.01$ &
$2.005$ \\
$\cos(i)$ &
cosine Earth inclination to spin axis &
$M,\cos(i)\sim N(\boldsymbol{\mu}^{\star},\boldsymbol{\Sigma}^{\star}) $ &
$0.0424_{-0.0030}^{+0.0030}$ &
$0.00$ &
$0.040$ \\
%\hline
&with joint prior PDF $N(\boldsymbol{\mu}^{\star},\boldsymbol{\Sigma}^{\star})$  & $\boldsymbol{\mu}^{\star}=[2.082,0.0427]^{\top}$\\
&&$\boldsymbol{\Sigma}^{\star}=
\begin{bmatrix}
0.0703^{2} & 0.0131^{2} \\
0.0131^{2} & 0.00304^{2}
\end{bmatrix}$\\
\hline
$R_{\textrm{eq}}$ $[$km$]$ &
coordinate equatorial radius &
$R_{\textrm{eq}}\sim U(3r_{\rm g}(1),16)$ &
$12.49_{-0.88}^{+1.28}$ &
$0.66$ &
$11.24$ \\
%\hline
&with compactness condition & $R_{\textrm{polar}}/r_{\rm g}(M)>3$\\
&with effective gravity condition & $13.7\leq \log_{10}g_{\textrm{eff}}(\theta)\leq15.0$,~$\forall\theta$\\
\hline
$\Theta_{p}$ $[$radians$]$ &
$p$ region center colatitude &
$\cos(\Theta_{p})\sim U(-1,1)$ &
$1.27_{-0.35}^{+0.48}$ &
$0.32$ &
$1.28$ \\
$\Theta_{s}$ $[$radians$]$ &
$s$ region center colatitude &
$\cos(\Theta_{s})\sim U(-1,1)$ &
$1.89_{-0.49}^{+0.36}$ &
$0.29$ &
$1.70$ \\
$\phi_{p}$ $[$cycles$]$ &
$p$ region initial phase &
$\phi_{p}\sim U(-0.5,0.5)$, wrapped &
bimodal &
$3.72$ &
$-0.255$\\
$\phi_{s}$ $[$cycles$]$ &
$s$ region initial phase &
$\phi_{s}\sim U(-0.5,0.5)$, wrapped &
bimodal &
$3.68$ &
$-0.323$ \\
$\zeta_{p}$ $[$radians$]$ &
$p$ region angular radius &
$\zeta_{p}\sim U(0,\pi/2)$ &
$0.115_{-0.030}^{+0.045}$ &
$2.72$ &
$0.151$ \\
$\zeta_{s}$ $[$radians$]$ &
$s$ region angular radius &
$\zeta_{s}\sim U(0,\pi/2)$ &
$0.115_{-0.030}^{+0.045}$ &
$2.70$ &
$0.124$ \\
&no region-exchange degeneracy & $\Theta_{s}\geq\Theta_{p}$\\
&nonoverlapping hot regions & function of $(\Theta_{p}, \Theta_{s}, \phi_{p}, \phi_{s}, \zeta_{p}, \zeta_{s})$\\
\hline
$\log_{10}\left(T_{p}\;[\textrm{K}]\right)$ &
$p$ region \TT{NSX} effective temperature &
$\log_{10}\left(T_{p}\right)\sim U(5.1,6.8)$, \TT{NSX} limits &
$6.022_{-0.046}^{+0.042}$ &
$3.20$ &
$6.054$ \\
$\log_{10}\left(T_{s}\;[\textrm{K}]\right)$ &
$s$ region \TT{NSX} effective temperature &
$\log_{10}\left(T_{s}\right)\sim U(5.1,6.8)$, \TT{NSX} limits &
$6.025_{-0.046}^{+0.042}$ &
$3.23$ &
$6.093$ \\
$D$ $[$kpc$]$ &
Earth distance &
$D\sim \texttt{skewnorm(1.7, 1.0, 0.23)}$ &
$1.20_{-0.15}^{+0.17}$ &
$0.07$ &
$1.57$ \\
$N_{\textrm{H}}$ $[10^{20}$cm$^{-2}]$ &
interstellar neutral H column density &
$N_{\textrm{H}}\sim U(0,10)$ &
$1.15_{-0.82}^{+1.50}$ &
$1.33$ &
$0.25$ \\
\hline
$\alpha_{\rm{NICER}}$ &
\NICER effective-area scaling &
$\alpha_{\rm{NICER}},\alpha_{\rm{XMM}}\sim N(\boldsymbol{\mu},\boldsymbol{\Sigma})$ &
$0.99_{-0.10}^{+0.10}$ &
$0.00$ &
$1.00$\\
$\alpha_{\rm{XMM}}$ &
\xmm effective-area scaling &
$\alpha_{\rm{NICER}},\alpha_{\rm{XMM}}\sim N(\boldsymbol{\mu},\boldsymbol{\Sigma})$ &
$0.99_{-0.10}^{+0.10}$ &
$0.02$ &
$0.89$ \\
&with joint prior PDF $N(\boldsymbol{\mu},\boldsymbol{\Sigma})$  & $\boldsymbol{\mu}=[1.0,1.0]^{\top}$\\
&&$\boldsymbol{\Sigma}=
\begin{bmatrix}
0.104^{2} & 0.100^{2} \\
0.100^{2} & 0.104^{2}
\end{bmatrix}$\\
\hline
\hline
&Sampling process information&&& \\
\hline
&number of free parameters: $15$ &&& \\
&number of processes (multimodes): $1$ &&& \\
&number of live points: $4\times10^{4}$ &&& \\
&SE:$^{\mathrm{a}}$ $0.01$ &&& \\
&termination condition: $0.1$ &&& \\
&evidence: $\widehat{\ln\mathcal{Z}}= -21889.99\pm0.02$ &&&\\
&number of core hours: $84,348$ &&& \\ %(1908+94.5)*24 + 4*24*126*3 = 84348
&likelihood evaluations: $66,823,871$ &&& \\
%&nested replacements: $1292538$ &&& \\
\enddata
\tablecomments{\ \
We show the prior PDFs, $68.3\,\%$ credible intervals around the median $\widehat{\textrm{CI}}_{68\%}$, Kullback–Leibler divergence $D_{\textrm{KL}}$ in bits representing the prior-to-posterior information gain, and the maximum likelihood nested sample $\widehat{\textrm{ML}}$ for all the parameters. 
Parameters for the primary hot region are denoted with a subscript $p$ and the parameters for the secondary hot region with a subscript $s$.
Note that the zero phase definition for $\phi_p$ and $\phi_s$ differs by 0.5 cycles.
Additional details of the parameter descriptions, the prior PDFs, and the sampling process information are given in the notes of Table 1 in \citetalias{Riley2021}.
\\
$^{\mathrm{a}}$ Called the inverse of the hypervolume expansion factor in \citetalias{Riley2021}. \\
}
\end{deluxetable*}

\section{Discussion}\label{sec:discussion}

\subsection{Updated PSR J0740+6620 Parameters and Comparison to Older Results}\label{sec:discussion_comparison_to_old}

Our new inferred mass and radius for \joh, using NICER data from 2018 September 21 to 2022 April 21 and joint modeling with \xmm, are $M = 2.073\pm 0.069 $ \msol (largely unchanged from the prior) and $R = 12.49_{-0.88}^{+1.28}$ km (68\% credible intervals). 
The 90\% credible interval for the radius is $R = 12.49_{-1.34}^{+2.26}$ km and the 95\% credible interval is $R = 12.49_{-1.53}^{+2.69}$ km.
The results continue to disfavor a very soft EOS, with $R < 10.96$ km excluded at the 97.5\% level and $R < 11.15$ km excluded at the 95\% level (the $X\%$ credible interval runs from the $(50 - X/2)\%$ to the $(50+X/2)\%$ quantile).
These lower limits are more constraining than the corresponding headline values reported in \citetalias{Riley2021} ($R < 10.71$ km excluded at the 97.5\% level and $R < 10.89$ km excluded at the 95\% level). 
Taken together with the fact that gravitational-wave observations favor smaller stars for intermediate-mass ($\sim 1.4$ \msol) NSs \citep{GW170817_TD1,GW170817_TD2} the two measurement techniques provide tight and complementary bounding constraints on EOS models. 
Tighter lower limits on the radius of such a high-mass pulsar should be more informative regarding, for example, the possible presence of quark matter in NS cores \citep[see, e.g.,][]{Annala2022}. 

Comparison to the older results is complicated by the fact that changes have been made to both the data set and the analysis method. 
The old headline results from \citetalias{Riley2021}, with a radius of $12.39_{-0.98}^{+1.30}$ km, were obtained using a larger effective-area scaling uncertainty than in this paper. 
Using compressed effective area scaling uncertainties (as used in this paper), and importance sampling the original results with a new prior, \citetalias{Riley2021} reported $12.71_{-0.96}^{+1.25}$ km as the radius. 
However, in this paper we also use more extensive sampler settings than before (see Section \ref{sec:posterior_computation}), and this also contributes to the differences in the new reported headline values.  
The lower limit for the radius credible intervals seems largely insensitive to the sampler settings, whereas the upper limit is more sensitive due---in large part---to flatness of the likelihood surface at high radii. 
While we are not able to formally prove convergence of our inferred radius we have investigated the factors driving the upper limit of the credible interval, and on the basis of the analyses carried out do not expect substantial further broadening (for details see Section \ref{sec:nicerXxmm}). 
We note that \citet{Dittmann2024} obtain an $\sim 0.5$ km higher upper limit for the radius (when limiting the radius to be below 16 km as we do); possible reasons for this are discussed in Section \ref{sec:discussion_comparison_to_dittmann}.

Inspecting the posterior distributions in Figure \ref{fig:other_posteriors}, one can see that most of the large-radius solutions (explored better by the HR run) are on average connected to smaller and hotter regions closer to the rotational poles.  
The pointy ends of the curved posteriors seen in the radius--colatitude and spot size--temperature planes can be challenging for samplers to explore, hence our use in this paper of more extensive \MultiNest settings and targeted (restricted prior) runs to explore this space. 
Moving from SE = 0.1 to SE = 0.01 resulted in more extensive sampling of this region likely causing the broadening of the radius credible interval.
Similar slow broadening has not yet been encountered in X-PSI analyses for other NICER sources, which are brighter, but it is clearly something to be alert to (although sensitivity to the sampler settings was also seen in the analysis of \jdbl in \citealt{Vinciguerra2024bravo}). 
Additional independent prior constraints on hot spot properties that might cut down the range of possibilities would be extremely helpful for \joh. 

Some of the model parameters are constrained more tightly with the new \NICER data set (while still consistent with the old results) in both the \NICER-only and the joint \NICER and \xmm analyses (mostly regardless of the sampler settings). 
In particular, the angular radii of the hot regions are now constrained around $35\,\%$ more tightly, and temperatures around $20\,\%$ more tightly. 
The inferred regions are on average slightly smaller and hotter than before.  
In addition, the interstellar hydrogen column density is constrained at least 20\,\% more tightly toward the lower end of the prior.
The credible intervals for the hot region colatitudes also become narrower by $10-15\,\%$, but only when using the same sampler settings. For the final HR joint \NICER and \xmm analysis the colatitude constraints do not significantly differ from the old results. 
The offset angle from an exactly antipodal hot region configuration is inferred to be $36\degr_{-11}^{+26}$, which is similar to the value of $38\degr_{-12}^{+23}$ inferred from the old analysis.
Thus, in both cases the offset angle is above $\sim 25\degr$ with $84\,\%$ probability (which is the same as reported in \citetalias{Salmi2022} using the 3C50 data set).  This implies a likely deviation from a centered-dipolar magnetic field.

\subsection{Scaling of Credible Intervals with Source to Background Count Ratio}\label{sec:discussion_scaling}

As shown in Section \ref{sec:results}, for identical sampler settings, we found that the constraints for the NS radius are roughly 20\% tighter when using the new NICER data set combined with the original \xmm data.  
However, when analyzing only the \NICER data, the radius constraints remained essentially unchanged. 
This can be attributed to the differences in the inferred ratio between source and background photons.

 This ratio is expected to influence the radius credible interval as outlined, for example, in Equation (4) of \citet{Psaltis2014}. This expression is obtained by assuming simplified model properties, e.g., considering small hot regions with isotropic emission (see also \citealt{PB06}), and assuming no uncertainty in the fundamental amplitude, background, and geometry parameters. Under these assumptions the uncertainty in the radius should scale as 
\begin{equation}
\frac{\Delta R_{\rm eq}}{R_{\rm eq}}\propto\frac{1}{\sqrt{N}}\left(\frac{S}{N}\right)^{-1}
\;,
\label{eq:radius_psaltis}
\end{equation}
where $S$ is the number of source counts, $N = S+B$ is the number of total counts, and $B$ is the number of background counts. Although this relation is extremely simplified, we can use it to make a rough comparison to the observed radius credible interval scaling.
Here we only compare runs performed with the same sampler settings (to avoid their influence).

We found that the fraction of inferred background photons increases with the new data set, especially in the case of the \NICER-only analysis (likely due to the different filtering of the data).
The expected number of source counts is even smaller for the new data than for the old data in a large fraction of samples; for the maximum likelihood sample in particular it is $20~\%$ smaller.
The old data allowed a large variety of different background versus source count solutions (see the green stepped curve in the left panel of Figure \ref{fig:bkg_inferred_data}), but for the new data the background is better constrained and higher on average (in comparison to the source counts). 
Thus, it is not trivial to predict how the radius credible intervals should change when increasing the number of observed counts. 

For the joint \NICER and \xmm analysis, the $S/N$ is better constrained, and does not change drastically when using the new \NICER data, as mentioned in Section \ref{sec:nicerXxmm}. 
The inferred $S$ for most samples in the joint analysis is higher with the new data, unlike for the \NICER-only analysis.
However, even in this case, accounting for the uncertainties in the inferred \NICER backgrounds would allow a significant variation in the predicted scaling for the radius credible intervals. 
Using $S$ and $S/N$ values based on one standard deviation limits of the inferred backgrounds,\footnote{Using 30 randomly drawn samples from the equally weighted posterior samples.} Equation \eqref{eq:radius_psaltis} predicts that credible intervals can become anything from $10~\%$ broader to $40~\%$ tighter.\footnote{Here we neglected the number of counts from \xmm as it is much smaller than that from \NICER.}
The observed $20~\%$ tightening is consistent with this.

Based on our results, it is not trivial to predict how the radius credible intervals evolve with more photons, at least in the presence of a large background and a large associated uncertainty.
Promisingly though, the inferred ratio between the source and background photons seems to be better constrained with the new data set. 
Thus, once this ratio is stable enough, the radius intervals are expected to tighten when increasing the total number of photons, as seen already in our joint \NICER and \xmm analysis.
Achieving a small $\sim \pm 5\,\%$ uncertainty for the \joh radius would still likely require increasing the current exposure time with \NICER to at least 8 Ms, based on the new headline results and the $1/\sqrt{N}$ dependence on the number of counts.

\subsection{Comparison to the Results of Dittmann et al. (2024)}\label{sec:discussion_comparison_to_dittmann}

Using the same \NICER and \xmm data sets as in this paper, an independent analysis carried out by \citet[][hereafter D24]{Dittmann2024} reports the \joh radius to be $12.92_{-1.13}^{+2.09}$ km.
This is broadly consistent with the $12.49_{-0.88}^{+1.28}$ km reported in this paper.
Although the radius credible interval of \citetalias{Dittmann2024} is about 50\% broader than here, the difference is notably smaller than the $\sim 80$\% difference between the headline results reported by \citetalias{Miller2021} and \citetalias{Riley2021} (the former using the same code as \citetalias{Dittmann2024}, the latter using X-PSI).
In particular, the radius lower limit---which as discussed previously is easier to constrain---is very similar for both, as was already the case in the 3C50 analyses of \citetalias{Salmi2022}.

The remaining differences between the results of this paper and \citetalias{Dittmann2024} could be related to different prior choices and/or different sampling procedures.\footnote{\citetalias{Dittmann2024} follow also the instrument channel choices of \citetalias{Miller2021}, but the effect of this was shown to be very small, see Figure \ref{fig:mr_posteriors_cross-calib}.}
When \citetalias{Dittmann2024} require radius to be less than $16$ km (the prior upper limit used in our analysis), they get $R=12.76_{-1.02}^{+1.49}$ km.
In addition, \citetalias{Dittmann2024} use a hybrid nested sampling (\MultiNest) and ensemble Markov Chain Monte Carlo (\textsc{emcee}, \citealt{emcee}) scheme, where a significant amount of the time can be spent in the ensemble sampling phase.
When \MultiNest is used, our resolution settings are higher in terms of the live points for runs reported in this paper. 
However, \MultiNest-only test comparisons were made using similar settings; when both teams use 4096 live points and SE = 0.01 the results are $12.70_{-0.97}^{+1.48}$ km for \citetalias{Dittmann2024} and $12.36_{-0.80}^{+1.06}$ km using X-PSI. 
Removing the samples with radius above 16 km the \citetalias{Dittmann2024} result becomes $12.66_{-0.94}^{+1.37}$ km.
As discussed in Appendix \ref{sec:sampling_efficiency} however, the effect of SE is connected to the initial prior volume and may thus not be directly comparable to \citetalias{Dittmann2024} who assume larger priors in many of the parameters (see \citetalias{Miller2021} and \citetalias{Riley2021} for details).
If we use the same number of live points (4096) as \citetalias{Dittmann2024} but drop the X-PSI SE to 0.0001 we get a radius of $12.55_{-0.92}^{+1.37}$ km, very close to the radius-limited \citetalias{Dittmann2024} \MultiNest-only result.\footnote{Note that our headline result used SE = 0.01 instead of SE = 0.0001 but also 40,000 live points instead of 4096, leading to roughly 10 times more accepted samples in the highest likelihood parameter space (also across other regions).}

In summary, the \MultiNest-only results of this paper and \citetalias{Dittmann2024} match fairly well when using the same radius prior upper limit and adjusting the sampler settings. 
Thus, the variation between the headline results may be related to different prior choices, to the differences between the \MultiNest and \textsc{emcee} samplers \citep{bilby}, and to the possible lack of convergence due to imperfect sampler settings. 

\section{Conclusions}
\label{sec:conclusions}

We have used a new \NICER data set, with about $1.13$ Ms additional exposure time, to reanalyze the pulse profiles of \joh.
Analyzing the data jointly with the same \xmm data as in \citetalias{Riley2021}, we infer the NS radius to be $12.49_{-0.88}^{+1.28}$ km, bounded by the $16\,\%$ and $84\,\%$ quantiles. 
This interval is consistent with the previous result and roughly equally wide. 
It was obtained with enhanced sampler settings, which were found to increase the radius upper limit compared to the result if using the old settings.
Even though we could not conclusively prove convergence of the upper limit, we consider the new results more robust after an extensive exploration of the likelihood surface and estimate the remaining systematic error to be $\lesssim 0.1$  km.
In addition, we found an expected parameter recovery in the analysis of simulated data resembling the new \joh observation (see Appendix \ref{sec:simulations}).
The radius lower limit is now also slightly more constraining than before; we exclude $R<11.15~\mathrm{km}$ at 95\% probability for this pulsar, and therefore the softest EOSs.

%\begin{acknowledgments}
\section*{Acknowledgments}
This work was supported in part by NASA through the \NICER mission and the Astrophysics Explorers Program. T.S., D.C., Y.K., S.V., and A.L.W. acknowledge support from ERC Consolidator grant No.~865768 AEONS (PI: Watts).  
The use of the national computer facilities in this research was subsidized by NWO Domain Science.
In addition, this work used the Dutch national e-infrastructure with the support of the SURF Cooperative using grant No. EINF-4664. 
Part of the work was carried out on the HELIOS cluster including dedicated nodes funded via the abovementioned ERC CoG.
Astrophysics research at the Naval Research Laboratory is supported 
by the NASA Astrophysics Explorer Program.
S.G. acknowledges the support of the Centre National d’Etudes Spatiales (CNES).
W.C.G.H. acknowledges support through grant 80NSSC23K0078 from NASA.
S.M. acknowledges support from NSERC Discovery grant RGPIN-2019-06077.
This research has made use of data products and software provided by the High Energy Astrophysics Science Archive Research Center (HEASARC), which is a service of the Astrophysics Science Division at NASA/GSFC and the High Energy Astrophysics Division of the Smithsonian Astrophysical Observatory.
We would also like to thank Will Handley, Johannes Buchner, Jason Farquhar, Cole Miller, and Alexander Dittmann for useful discussions. 
%\end{acknowledgments}

\facilities{\NICER, XMM}

\software{Cython \citep{Behnel2011}, GetDist \citep{Lewis2019}, GNU Scientific Library \citep{Galassi2009}, HEASoft \citep{heasoft2014}, Matplotlib \citep{Hunter2007}, MPI for Python \citep{Dalcin2008}, \MultiNest \citep{multinest09}, nestcheck \citep{Higson2018JOSS}, NumPy \citep{Walt2011},\PyMultiNest \citep{PyMultiNest}, Python/C language \citep{Oliphant2007}, SciPy \citep{Jones}, and \XPSI \citep{xpsi}.}

\bibliographystyle{aasjournal}
\bibliography{allbib}

\clearpage 
\appendix
%\figsetstart
%\figsetnum{2}
%\figsettitle{Posteriors other than space-time parameters}

%\figsetgrpstart
%\figsetgrpnum{7.1}
%\figsetgrptitle{Backgrounds for "synt1"}
%\figsetplot{f7_1.png}
%\figsetgrpnote{Inferred backgrounds for the synthetic data set “synt1”. See the caption in the main text for more details.}
%\figsetgrpend

%\figsetgrpstart
%\figsetgrpnum{7.2}
%\figsetgrptitle{Backgrounds for "synt2"}
%\figsetplot{f7_2.png}
%\figsetgrpnote{Inferred backgrounds for the synthetic data set “synt2”. See the caption in the main text for more details.}
%\figsetgrpend

%\figsetgrpstart
%\figsetgrpnum{7.3}
%\figsetgrptitle{Backgrounds for "synt3"}
%\figsetplot{f7_3.png}
%\figsetgrpnote{Inferred backgrounds for the synthetic data set “synt3”. See the caption in the main text for more details.}
%\figsetgrpend

%\figsetend

{
    \begin{figure}[t!]
    \centering
    \resizebox{\hsize}{!}{\includegraphics[
    width=\textwidth]
    {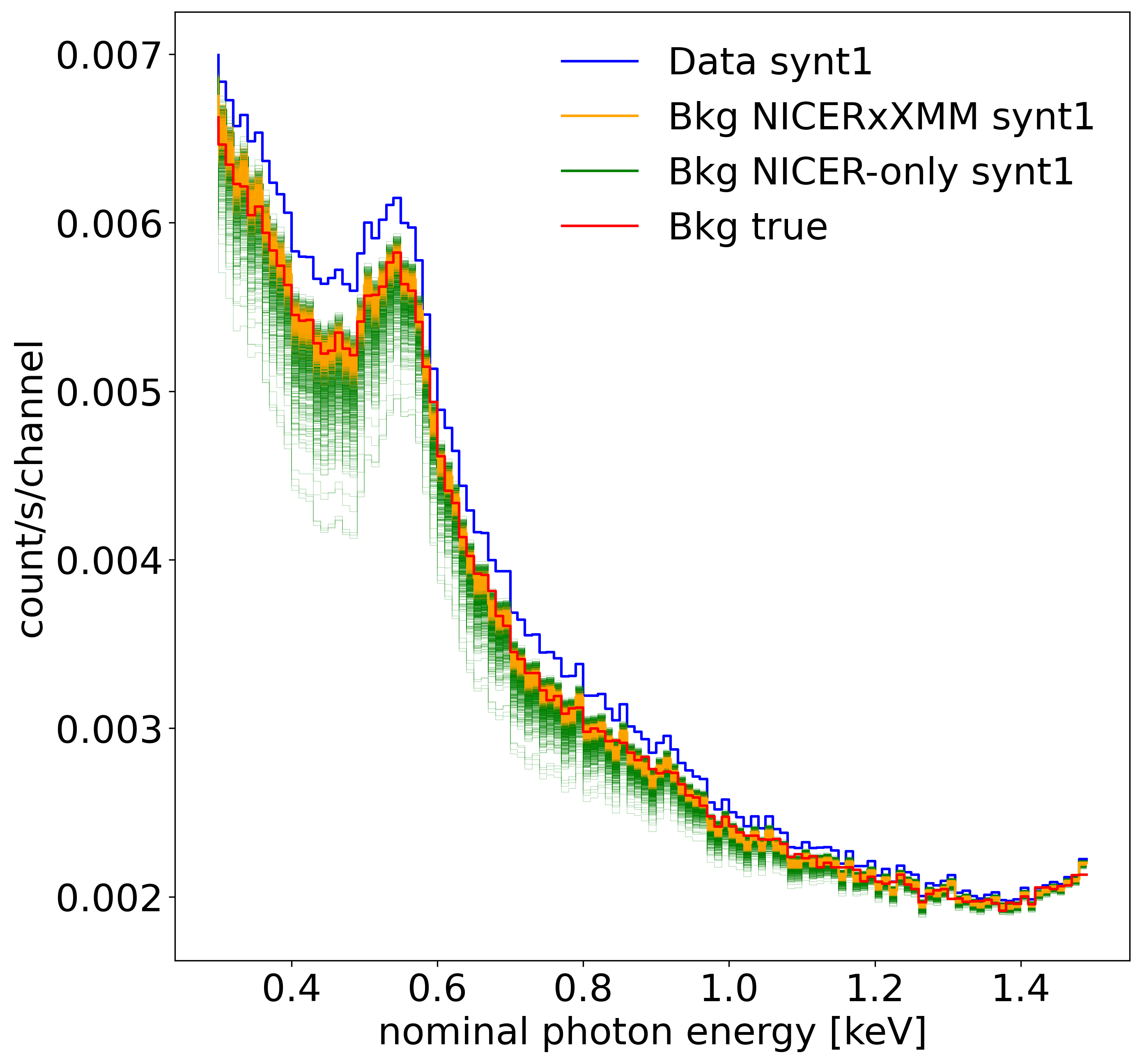}}
    \caption{\small{
     Comparison of the inferred \NICER background for the synthetic data set “synt1” based on either the \NICER-only or joint \NICER and \xmm analysis. 
     The blue stepped curve shows the synthetic data.
     The orange stepped curves show background curves that maximize the likelihood for 1000 equally weighted posterior samples from the joint \NICER and \xmm run. 
     The green stepped curves (partly hidden by the orange) show the same for samples from the \NICER-only run.
     The red stepped curve shows the injected background for the synthetic data.
     The complete figure set (three images), including also the inferred backgrounds for two other synthetic data realizations, is available in the online journal (HTML version).
    }}
    \label{fig:bkg_synthetic}
    \end{figure}
}

{
    \begin{figure*}[t!]
    \centering
    \includegraphics[
    width=\textwidth]{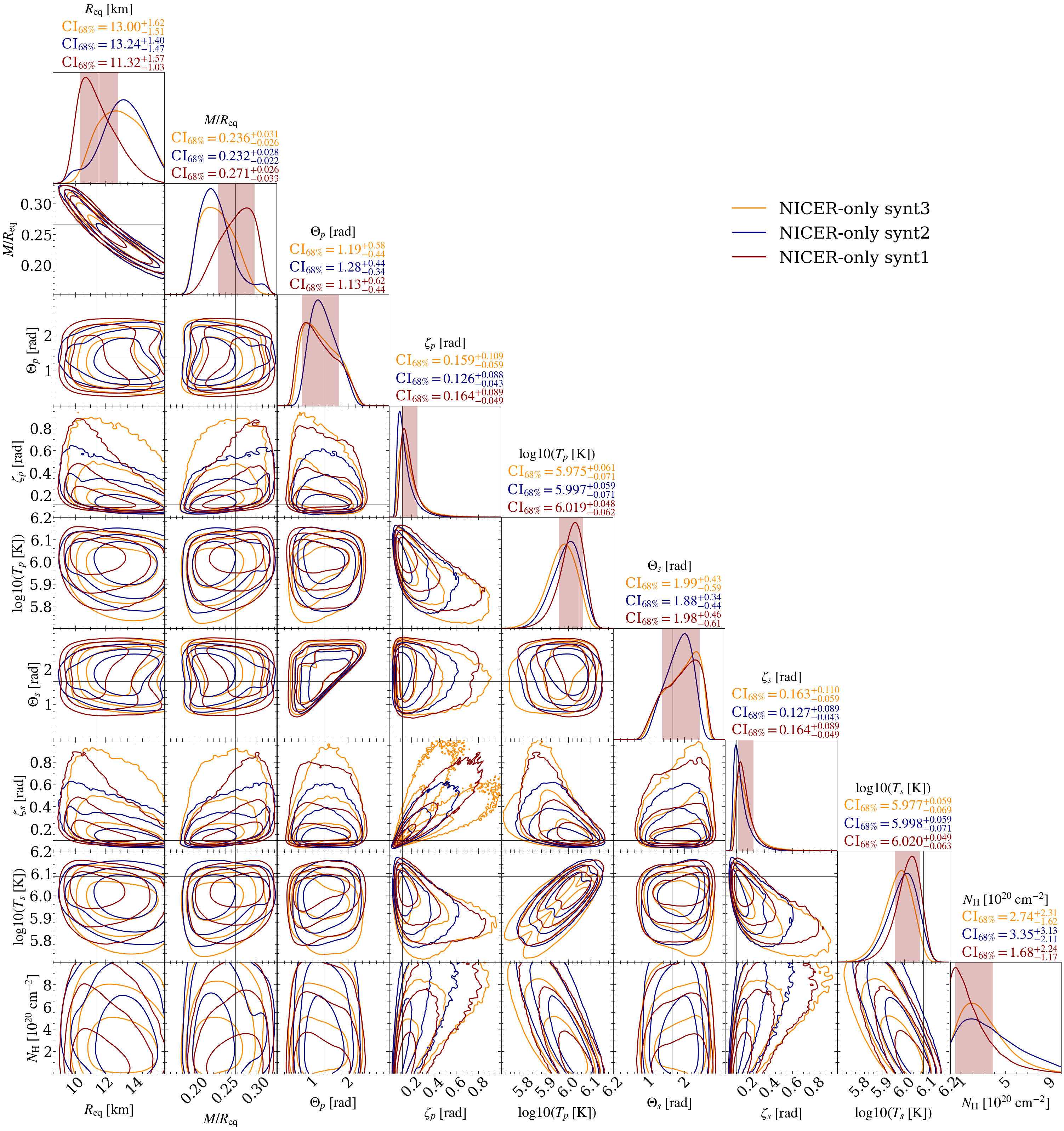}
    \caption{\small{
    Posterior distributions for the most run-to-run varying parameters using the synthetic \NICER data sets and the \texttt{ST-U} model.
    The shaded vertical bands show the $68.3\%$ credible intervals for the run with synthetic data labeled as “synt1.”
    The thin black lines represent the injected values. 
    See the caption of Figure \ref{fig:mr_posteriors} for more details about the figure elements.
    }}
    \label{fig:posteriors_synthetic}
    \end{figure*}
}

{
    \begin{figure*}[t!]
    \centering
    \includegraphics[
    width=\textwidth]{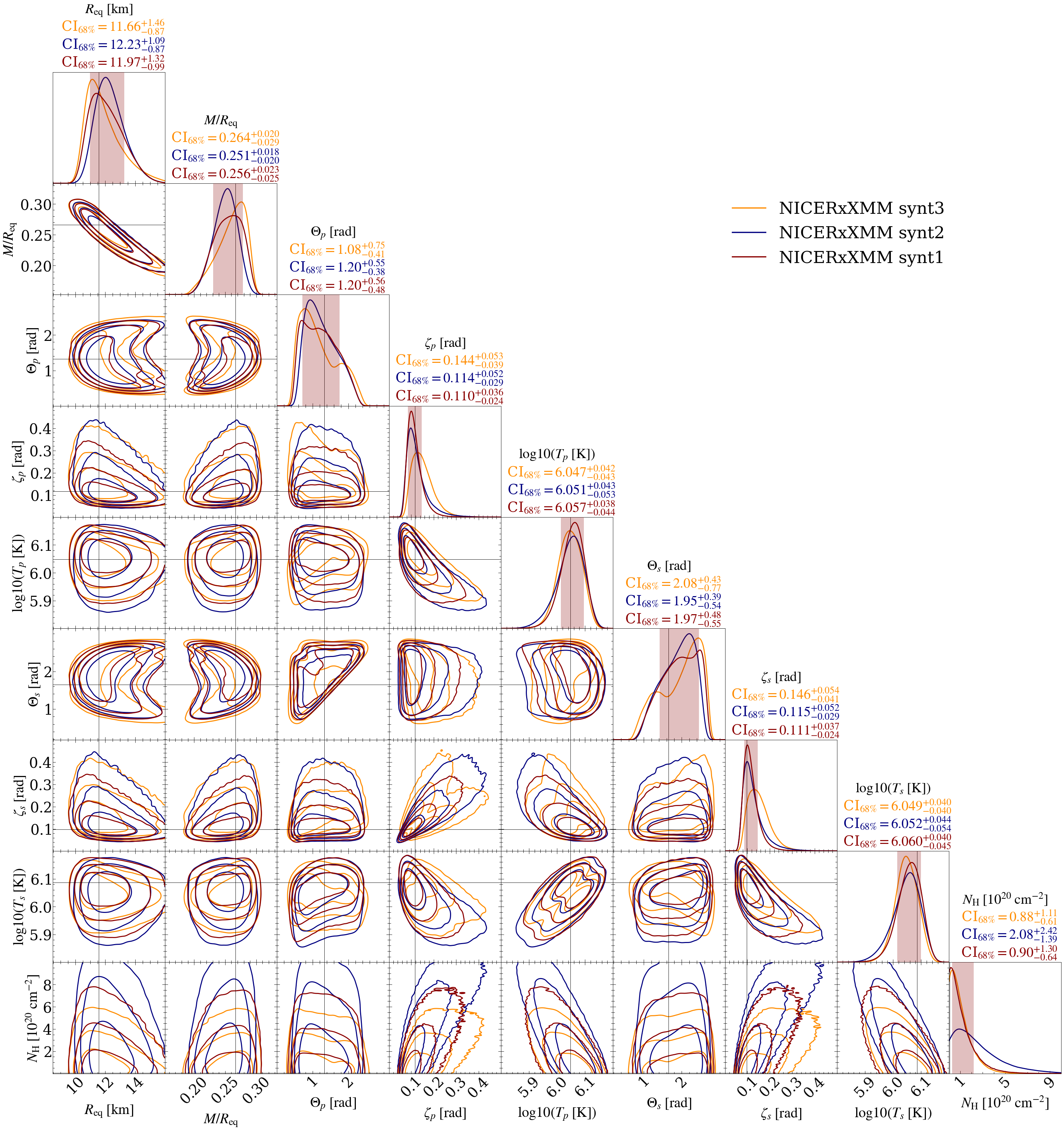}
    \caption{\small{
    Posterior distributions for the most run-to-run varying parameters using the synthetic \NICER and \xmm data sets and the \texttt{ST-U} model.
    The shaded vertical bands show the $68.3\%$ credible intervals for the run with synthetic data labeled as “synt1.”
    The thin black lines represent the injected values. 
    See the caption of Figure \ref{fig:mr_posteriors} for more details about the figure elements.
    }}
    \label{fig:posteriors_synthetic2}
    \end{figure*}
}

\begin{deluxetable}{ll}[b]
\tablecaption{Injected Model Parameters}
\tablehead{\multicolumn{1}{l}{Parameter} & \multicolumn{1}{l}{Injected Value}}
\startdata
$M$ $[\textit{M}_{\odot}]$ &
$2.088$ \\
$R_{\textrm{eq}}$ $[$km$]$ &
$11.57$ \\
$\Theta_{p}$ $[$radians$]$ &
$1.324$  \\
$\Theta_{s}$ $[$radians$]$ &
$1.649$ \\
$\phi_{p}$ $[$cycles$]$ &
$-0.258$  \\
$\phi_{s}$ $[$cycles$]$ &
$-0.328$\\
$\zeta_{p}$ $[$radians$]$ &
$0.117$   \\
$\zeta_{s}$ $[$radians$]$ &
$0.098$ \\
$\log_{10}\left(T_{p}\;[\textrm{K}]\right)$ &
$6.048$ \\
$\log_{10}\left(T_{s}\;[\textrm{K}]\right)$ &
$6.086$ \\
$\cos(i)$ &
$0.041$ \\
$D$ $[$kpc$]$ &
$1.187$ \\
$N_{\textrm{H}}$ $[10^{20}$cm$^{-2}]$ &
$0.067$ \\
$\alpha_{\rm{NICER}}$ &
$0.905$ \\
$\alpha_{\rm{XMM}}$ &
$0.811$ \\
\enddata
\tablecomments{\ \\  See the parameter descriptions in Table \ref{table:results}.}
\end{deluxetable}\label{table:true_values}

\section{Simulations}\label{sec:simulations}

To test the robustness of the analyses of this paper, we performed several inference runs using synthetic data.
We generated three different noise realizations for both the synthetic \NICER and \xmm data using the maximum likelihood parameter vector found in the initial new joint \NICER and \xmm analysis with real data (the run using the same sampler settings as in \citetalias{Riley2021}).
The parameters are shown in Table \ref{table:true_values}.
The input background spectrum for each instrument was set to be the one that maximized the instrument-specific likelihood for the real data (e.g., for \NICER the red curve in Figure \ref{fig:bkg_synthetic}).
For each synthetic data set, Poisson fluctuations were added to the sum of counts from the hot spots and the background using a different seed for the random number generator. 
The exposure times were matched to those of the true observations.

For the inference runs, we applied the same prior distributions as in the analysis of the real data.
For the \MultiNest resolution settings we used the same choices as for the real data, but with SE = 0.1 to keep the computational cost manageable. 
We note that the credible interval for the radius and a couple of other parameters was larger for the headline results with SE = 0.01. 
However, the expected parameter recovery found from the simulations still indicates that at least the headline intervals are unlikely to be heavily underpredicted.

We performed six inference runs in total: three analyzing only \NICER data (one for each noise realization) and three analyzing the \NICER and \xmm data jointly (one for each data pair with the same seed when creating the data).
The results of these runs are shown in Figures \ref{fig:posteriors_synthetic} (\NICER only) and \ref{fig:posteriors_synthetic2} (joint \NICER and \xmm) for the most varying parameters (in contrast, the posteriors for the mass, inclination, and distance were always found to follow closely their priors).
We see that the true radius, compactness, hot spot properties, and the hydrogen column density $N_{\mathrm{H}}$ values are better recovered when jointly fitting \NICER and \xmm in all the three cases. 
However, even for the \NICER-only analysis, the injected radius is found within the $68~\%$ credible limits in two out of three cases (the expectation being between one and three).\footnote{We estimate the expected ranges based on the sample size and $16~\%$ and $84~\%$ quantiles of the percent point function of the binomial distribution with $68~\%$ success rate. When considering many model parameters combined, the range is only indicative since it assumes independence between the parameters, which are instead correlated.} 
If accounting for all the sampled parameters in the \NICER-only runs, the injected values are found within the $68~\%$ intervals in $67~\%$ of the cases (the expectation being between $62~\%$ and $76~\%$).
For the joint \NICER and \xmm analyses, all the injected radii, and $78\,\%$ of all the injected parameters, are found inside the $68~\%$ interval (the expectations being the same as for the \NICER-only case). 
The inferred background spectra were also found to resemble the true background for all the runs, although with less scatter in the case of the joint \NICER and \xmm runs, as seen in Figure \ref{fig:bkg_synthetic}.

We can also see that the inferred credible intervals for the simulated data are a bit larger than for the corresponding true data (see Section \ref{sec:results} for the true data).
In the case of the \NICER-only analysis, the width of the $68~\%$ radius interval is around $2.5-3.1$ km for all simulations, and about $1.9$ km for the true data.
In the case of the joint \NICER and \xmm analysis, the width of the same interval is around $1.9-2.4$ km for the simulations, and about $1.8$ km for the true data with the same sampler settings (but $2.2$ km for the higher-resolution headline results).
However, this difference is likely due to the small number of runs on the simulated data.
We also performed two additional low-resolution \NICER-only runs (with $4\times10^{3}$ live points instead of $4\times10^{4}$ used otherwise) to see how the credible intervals for synthetic data depend on the sampling resolution.
We found 0.3--0.4 km wider $68\%$ intervals for the radius when using the higher resolution, which is in accordance with what was found for the true data in \citetalias{Riley2021}.

In the current analysis, we did not test how the observed parameter recovery could change if selecting other injected values.
More simulation tests with \XPSI were recently reported by \citet{Kini2023a} and \citet{vinciguerra2023sim} with different parameters, models, and instruments, showing the expected recovery when the data were created and fitted with the same model.
Our results with the \joh-like simulated data support those findings but show also that analyzing many instruments jointly may improve the accuracy of the recovered values, given our assumptions of cross-calibration uncertainties.
\section{Treatment of Sampling Efficiency in X-PSI}\label{sec:sampling_efficiency}

To clarify the terminology and treatment of the \MultiNest SE parameter in X-PSI, we recap here the procedure described in Appendix B.5.3 of \citet{riley_thesis}.
As mentioned there (and in Footnote \ref{footnote:SE}), the native SE setting is modified to account for the initial prior volume, which can differ from unity in our models unlike in the original \MultiNest  algorithm \citep[see Algorithm 1 in][]{multinest09}.
That algorithm uses a prior volume that is shrunk at each iteration so that the remaining expected prior volume is $\exp(-i/N)$, where $i$ is the iteration number and $N$ is the number of live points. 
This is used to set the minimum volume $V_{\mathrm{m}}$ of the approximate isolikelihood-bounding ellipsoid from which higher-likelihood points are drawn at each iteration.
To avoid sampling from a too small volume (in case the ellipsoidal approximation is not accurate), the minimum volume is additionally enlarged to $V_{\mathrm{m}}=e\exp(-i/N)$, where $e$ is the expansion factor and the inverse of SE.
This means, in practice, that $V_{\mathrm{m}}$ shrinks exponentially at every iteration, but the initial value is now $e$ instead of $1$.

In X-PSI analyses, however, the initial prior volume is usually less than $1$ due to the rejection rules applied. 
For example, if the star is too compact, a likelihood value below the \MultiNest  \texttt{log\_zero} threshold is returned so that the sample will be automatically ignored.
Therefore, we set the shrinkage of $V_{\mathrm{m}}$ to start from $He$ (instead of $e$), where $H$ is the true estimated initial prior volume.
This is done by setting the nominal input value of the SE parameter to $1/(He)$ in the code. 
When reporting SE values we still refer to $1/e$, since that describes the enlargement relative to the true initial prior.

\end{document}